\documentclass[manuscript,screen]{acmart} % removed "review" option: it adds margin line numbers, which arXiv disallows

\AtBeginDocument{%
  }

\usepackage{algorithmic}
\usepackage{graphicx}
\usepackage{textcomp}
\usepackage{xcolor}
\usepackage{caption}
\usepackage{listings}
\usepackage{listingsutf8}
\usepackage{verbatim}

\usepackage{booktabs}
\usepackage{tabularx}
\usepackage{ragged2e} % for \RaggedRight
\newcolumntype{Y}{>{\RaggedRight\arraybackslash}X}

\lstdefinestyle{mypython}{
  language=Python,
  inputencoding=utf8,
  columns=fullflexible, % important for wide CJK glyphs
  basicstyle=\ttfamily\scriptsize,
  keywordstyle=\color{blue},
  stringstyle=\itshape,
  commentstyle=\color{gray}\itshape,
  showstringspaces=false,
  breaklines=true,
  frame=none,
  numbers=left,
  numberstyle=\tiny\color{gray},
}

\lstdefinestyle{myjava}{
  language=Java,
  basicstyle=\ttfamily\scriptsize, % reduce font size
  numbers=left,                       % line numbers (optional)
  numberstyle=\tiny,
  stepnumber=1,
  numbersep=5pt,
  tabsize=2,
  showstringspaces=false,
  breaklines=true,
  keywordstyle=\color{blue}\bfseries,
  commentstyle=\color{teal},
  stringstyle=\color{red},
  morekeywords={var} % add Java 10+ keywords if needed
}

\usepackage{xspace}

\usepackage{booktabs}
\usepackage{makecell}
\usepackage{tabularx}

\newcommand{\darkred}{\color[RGB]{139,0,0}}
\newcommand{\darkgreen}{\color[RGB]{0,100,0}}
\definecolor{darkgreen}{rgb}{0.0, 0.5, 0.0}
\newcommand{\ie}{\emph{i.e.,}\xspace}
\newcommand{\eg}{\emph{e.g.,}\xspace}

\newcommand{\etal}{\emph{et~al.}\xspace}
\newcommand{\secref}[1]{Section~\ref{#1}\xspace}

\newcommand{\figref}[1]{Fig.~\ref{#1}\xspace}

\newcommand{\tabref}[1]{Table~\ref{#1}\xspace}

\newcommand{\gpt}{GPT-4o \emph{mini}\xspace}
\newcommand{\ds}{DeepSeek\xspace}
\newcommand{\cl}{Claude\xspace}
\newcommand{\gp}{GPT\xspace}

\newcommand{\sonar}{\textsc{SonarCloud}\xspace}

\newcommand{\nloc}{NLOC\xspace}
\newcommand{\cc}{CC\xspace}
\newcommand{\cmt}{Cmt\xspace}
\newcommand{\cmd}{\% Cmt\xspace}
\newcommand{\cogn}{Cogn\xspace}
\newcommand{\smells}{Smells\xspace}
\newcommand{\warnings}{Warnings\xspace}
\newcommand{\bugs}{Bugs\xspace}
\newcommand{\pmd}{\textsc{PMD}\xspace}
\newcommand{\pylint}{\textsc{pylint}\xspace}
\newcommand{\flake}{\textsc{flake8}\xspace}

\newcommand\nb[2]{\textbf{[#1: #2]}}

\newcommand\MAX[1]{\textcolor{green}{\nb{MAX}{#1}}}

\newcommand\BOWEN[1]{\textcolor{brown}{\nb{BOWEN}{#1}}}

\newcommand{\rev}[1]{\textcolor{black}{#1}}

\newcommand{\revminor}[1]{\textcolor{black}{#1}}

\usepackage[most]{tcolorbox}
\newtcolorbox{promptbox}{colback=white, arc=0.5mm, top=1mm, bottom=1mm, left=1mm, right=1mm, title=System prompt used for generation}

\newtcolorbox{resultbox}{colback=white, arc=0.5mm, top=1mm, bottom=1mm, left=1mm, right=1mm}

\begin{document}

%\title{Large Language Models for Multilingual Code Generation: A Curated Benchmark and a Study on Code Quality}
\title{Large Language Models for Code Generation \rev{from Multilingual Prompts}: A Curated Benchmark and a Study on Code Quality}

\author{Saima Afrin}
\orcid{0009-0008-4106-6838}
\affiliation{
  \institution{William \& Mary}
  \city{Williamsburg}
  \country{VA, USA}}
\email{safrin@wm.edu}

\author{Alessandro Midolo}
\orcid{0000-0002-9575-8054}
\affiliation{
  \institution{University of Catania}
  \city{Catania}
  \country{Italy}}
\email{alessandro.midolo@unict.it}

% \author{Saima Afrin}
% \orcid{0009-0008-4106-6838}
% \affiliation{
%   \institution{William \& Mary}
%   \city{Williamsburg}
%   \country{VA, USA}}
% \email{safrin@wm.edu}

\author{Camilo Escobar-Velásquez}
\orcid{0000-0001-8414-9301}
\affiliation{
  \institution{Universidad de los Andes}
  \city{Bogota}
  \country{Colombia}}
\email{ca.escobar2434@uniandes.edu.co}

\author{Mario Linares-Vásquez}
\orcid{0000-0003-0161-2888}
\affiliation{
  \institution{Universidad de los Andes}
  \city{Bogota}
  \country{Colombia}}
\email{m.linaresv@uniandes.edu.co}

\author{Weiyuan Ding}
\orcid{}
\affiliation{
  \institution{North Carolina State University}
  \city{Raleigh}
  \country{USA}}
\email{wding8@ncsu.edu}

\author{Bowen Xu}
\orcid{0000-0002-1006-8493}
\affiliation{
  \institution{North Carolina State University}
  \city{Raleigh}
  \country{USA}}
\email{bxu22@ncsu.edu}

\author{Massimiliano Di Penta}
\orcid{0000-0002-0340-9747}
\affiliation{
  \institution{University of Sannio}
  \city{Benevento}
  \country{Italy}}
\email{dipenta@unisannio.it}

\author{Antonio Mastropaolo}
\orcid{0000-0002-7965-7712}
\affiliation{
  \institution{William \& Mary}
  \city{Williamsburg}
  \country{VA, USA}}
\email{amastropaolo@wm.edu}

\begin{abstract}
Large Language Models (LLMs) perform differently on identical tasks when prompted in different languages—a phenomenon known as language bias. Although well-documented for general text generation, the extent to which language bias affects code generation quality and programming conventions remains largely unexplored.
To address this gap, we investigate how the natural language used to describe programming tasks influences the quality of source code generated by three prominent LLMs: \rev{\gpt}, \ds, and \cl. Our study includes 460 coding tasks spanning Python (230 tasks) and Java (230 tasks). We \rev{translate and manually curate} the original English prompts into four diverse languages: Chinese, Hindi, Spanish, and Italian, ensuring linguistic accuracy while preserving technical meaning. We evaluate the resulting code across multiple dimensions: functional correctness through test passage rates, structural quality via established code metrics, potential issues identified by static analysis tools, and lexical characteristics, including the natural language used in identifiers and comments. Results indicate that (i) source code generated from the English queries is not necessarily better in terms of passed test and quality metrics, (ii) the quality for different languages varies depending on the programming language and LLM being used, and (iii) the generated code tends to contain mixes of comments and literals written in English and the prompt language.
\end{abstract}




\maketitle
\sloppy

\section{Introduction} \label{sec:introduction}

%\BOWEN{I think vibe coding phenomenon (\url{https://en.wikipedia.org/wiki/Vibe_coding}) could potentially be a part of motivation on why we should care about prompt in different languages.}

Large Language Models (LLMs) have transformed software engineering automation by raising the bar to unprecedented levels, enabling practitioners to automate various coding tasks such as code generation \cite{fakhoury2024llm, jiang2024surveylargelanguagemodels, chen2024survey}, summarization \cite{szalontai2024large, sun2024source, zhang2024review}, program repair \cite{fan2023automated, pearce2023examining, cao2025study}, and testing \cite{wang2024software, xiong2023program}--among other \cite{liu2022few, ahmed2024automatic, sun:icsme2021, ryan2024code}. %\ANTONIO{Saima, add LLM-based inference only papers}
%such as automated bug fixing, program repair, code summarization, code completion, and software testing. 
%Nowadays, developers working around the world interact more and more frequently with LLM-based assistants during their everyday tasks or leverage tools within their Integrated Development Environments (IDEs), such as GitHub Copilot \cite{copilot} or Gemini \cite{gemini} to support code. Even there are already literature efforts devoted to comprehensively reviewing software engineering (SE) activities that have been automated using LLMs from the research perspective \cite{watson:tosem2022, hou2023large}.  There is no doubt of the potential and latent benefits of using LLMs to support software engineering, and as a consequence, different aspects of their impact, capabilities, and implementation must be investigated.
The widespread adoption of commercial solutions such as GitHub Copilot \cite{imai:icse2022, doderlein2022piloting, pudari2023copilot, yetiştiren2023evaluatingcodequalityaiassisted, Mastropaolo:icse2023} and ChatGPT \cite{rajbhoj2024accelerating, ahmad2023towards, khoury2023secure, sobo2025evaluating}, which act as AI companions for developers, demonstrates the transformative power of LLMs in software engineering. However, this success comes with underlying challenges rooted in how these models are trained. LLMs learn by processing vast quantities of unstructured data from the wild, including source code, comments, documentation, and natural language descriptions. While this broad training approach enables learning from diverse sources, it introduces significant downsides.%As shown in previous work \cite{pepe2024hugging}, such bias can be of different types, including---according to a taxonomy by Mehrabi \etal \cite{MehrabiMSLG21}, inappropriate content bias, historical bias, population bias, or social bias.
%Among the various types of bias, \emph{population bias}, and, specifically, language bias is particularly significant for LLM-based code recommenders, as these models are utilized globally, and the language used to interact with them can vary across different user demographics. In this regard, 

%the training data is likely to exhibit certain language distributions, with English being a \emph{lingua franca} for writing source code elements (\ie identifiers, comments, literal) and related descriptions--if any. Nonetheless, forges like GitHub contain repositories whether other languages are used, though these are not the majority~\cite{prana2019categorizing}. 

Let us consider the following scenario. A developer, working in a country where English is not the main language, receives a specification written in their language. To seek assistance, the developer interacts, using their mother tongue, with an LLM-based assistant.
When the query is posed in a language other than English, however, the output of the LLM may differ. This variation can arise from distribution shifts in the training data, since non-English content may be less prevalent or unevenly represented in the model's training corpus.

%the developer may question whether the LLM accurately understood the request and how well the generated code meets the intended requirements. 
Inspired by this scenario, this paper aims to address the following question:

\begin{tcolorbox}[colback=blue!5!white, colframe=blue!50!black, title=\textbf{Research Question}]
\centering
\emph{How does the use of different languages in coding prompts affect the quality of generated source code?}
\end{tcolorbox}

Existing research \cite{asare2024user,asare2023github,kharma2025security,perry2023users,sandoval2023lost,yetiştiren2023evaluatingcodequalityaiassisted} has examined the quality of source code generated by LLMs, primarily focusing on correctness---which measures the ability of the model to generate code that passes test cases---and overall code quality. A previous preliminary study showed correctness differences in a programming contest~\cite{koynagi24}, while another study has considered, on automatically-translated tasks from the HumanEval\cite{humaneval} benchmark, the extent to which generated code from multilingual prompts passes different proportions of tests~\cite{humanevalxl2024peng}.
%to solely assess the pass test of generated code among different languages of prompting and different programming languages.

% there is no systematic analysis of the effect of prompt language on the quality of generated programs.

%However, to the best of our knowledge, no research has explored the extent to which prompts written in different languages may influence the quality of generated code, potentially introducing biases in code generation.

Our work studies bias in code generation from multilingual prompts considering three key additional aspects not dealt by previous work: (i) manually curated prompt translation (necessary to make prompts looking natural), (ii) quality seen not only in terms of passed test, but also using quality metrics and static analysis warnings, and (iii) the use of more complex tasks, from the CoderEval~\cite{hao2024codereval} benchmark.

For prompts, we consider four languages other than English: Chinese, Hindi, Spanish, and Italian. The selection of these languages is based on their global prominence and linguistic diversity, ensuring a broad representation of writing systems.
Chinese, one of the most widely spoken languages worldwide, employs a \emph{logographic} writing system, which contrasts with the alphabetic systems used by the other selected languages (\ie Italian and Spanish). Hindi, a major language in South Asia, is written in the \emph{Devanagari} script and features a distinct linguistic and syntactic structure.
Finally, Spanish and Italian represent Latin languages, with Spanish being an official language in 22 countries \cite{languages}, and Italian representing a more geographically concentrated speaking population.
%Italian, on the other hand, was chosen to examine how LLMs handle a Romance language with a more geographically concentrated speaker population, providing a useful contrast to the widespread distribution of Spanish.
%This selection of languages enables us to (i) examine potential bias in LLM-based generated code and (ii) assess how prompt language influences code quality and generation outcomes.

%\MAX{we need to explain why these languages}
To conduct the study, we took 460 code generation tasks from the CoderEval~\cite{hao2024codereval} benchmark and considered two popular programming languages, \ie Python and Java. We began by creating code generation prompts from the benchmark queries--originally written in English--into the other four languages. In contrast to recent studies~\cite{humanevalxl2024peng}, the translated prompts were manually validated by native-speaker authors to minimize the risk of (mis)-translation and ensure the original intent of the prompts was accurately preserved. Then, we generated code with three state-of-the-art LLMs, \ie  \gpt \cite{gpt4}, \ds \cite{deepseekai2025report}, and \cl \cite{AnthropicClaude3}. 

Once the code has been generated, we performed a quality assessment over several dimensions, including (i) code correctness, (ii) code quality indicators, including various types of source code metrics, and (iii) static analysis tool warnings.
Our results indicate that source code generated with prompts other than English is not necessarily less correct, yet this depends on the programming language and on the language itself. Metrics and static analysis tools do not indicate lower quality when English is not used, yet they show how the code is structurally different. Finally, while LLMs tend to generate identifiers in English, comments and string literals show a high variability, potentially impacting long-term maintainability.
%Such results have implications for developers who use LLMs to generate code starting from specifications written in their own languages. 

Overall, we summarize the contributions of our empirical study as follows:

\begin{itemize}
    \item Our work contributes with a large, multi-lingual dataset comprising (i) the queries with manually-curated translation into four languages, (ii) the generated code with three LLMs, and (iii) the code quality indicators along the dimensions mentioned above, \ie test results, metrics, static analysis warnings, and language used for the source code lexicon.
    \item We derive several insightful findings from the experimental results and their implications for both practitioners and educators in the SE community, including:
        \begin{itemize}
        \item Counterintuitively, prompting in English does not always lead to the best correctness in code generation tasks. \revminor{Prompting in Chinese leads to consistently better performance in Python code generation on CoderEval; however, this advantage does not fully generalize to class-level tasks (ClassEval), and no language dominates across Java.}
        
        %\item Static analysis warnings and metrics reveal that non-English prompts do not necessarily reduce code quality. Non-English prompts often yield simpler code but more style/convention violations, with Hindi showing riskier outputs, whereas Chinese, Spanish, and Italian generally improve or remain stable. 
        \item \revminor{Across the size, complexity, and comment metrics and the static-analysis warnings from \flake, \pylint, \pmd, and \sonar examined in this study, non-English prompts do not necessarily reduce code quality.} Non-English prompts often yield simpler code but more style/convention violations, with Hindi showing riskier outputs, whereas Chinese, Spanish, and Italian generally improve or remain stable.
        
        \item LLMs show notable inconsistency in generating literals and comments, often defaulting to English or mixing languages within the same code. While function signature language partially drives lexicon, consistent adherence to the target language is not guaranteed.  
    \end{itemize}
\end{itemize}

The remainder of the article is organized as follows. Section~\ref{sec:related} presents the essential background and related work. Section~\ref{sec:design} describes the overall design of our study. Section~\ref{sec:results} reports and analyzes the main findings. Section~\ref{sec:implications} discusses the broader implications and potential applications of our work. Section~\ref{sec:threats} outlines the threats to validity and their possible impact. Finally, Section~\ref{sec:conclusion} concludes the paper and suggests directions for future research.

% !TEX root = ../main.tex
\section{Background and Related Work}
\label{sec:related}

In this section, we highlight related literature concerning: (i) the quality of LLM-generated code and, (ii) multilingual inconsistencies in LLMs.

\subsection{Quality of LLMs-based Generated Code}
\label{sec:quality}
Extensive research has been conducted to evaluate the quality and reliability of code generated by LLMs. Perry \etal \cite{perry2023users} found that LLM-generated code contains more security vulnerabilities than human-written code.  Similarly, Sandoval \etal \cite{sandoval2023lost} identified a 10\% increase in security vulnerabilities in LLM-generated C code.
%further highlighting concerns regarding the robustness of AI-assisted coding.
Asare \etal \cite{asare2024user} found that while GitHub Copilot improved security in complex programming tasks, its impact on simpler problems was minimal. 
Yetiştiren \etal \cite{yetiştiren2023evaluatingcodequalityaiassisted} evaluated ChatGPT \cite{gpt4}, GitHub Copilot \cite{copilot}, and Amazon CodeWhisperer \cite{codewhisperer}, using SonarQube \cite{SonarQube2024} to assess security, maintainability, and reliability.
Further, Asare \etal \cite{asare2023github} reported that Copilot replicated security vulnerabilities in 33.3\% of cases but mitigated 25.5\%. Khoury \etal \cite{khoury2023secure} examined ChatGPT's security performance across programming languages, showing frequent failures to meet security standards.%, though iterative prompting improved results.
Mastropaolo \etal \cite{Mastropaolo:icse2023} revealed GitHub Copilot's inability to handle semantically equivalent natural language code descriptions, generating code recommendations in only 46\% of cases and exhibiting a 28\% correctness gap between variant descriptions.
Kharma \etal \cite{kharma2025security} explored the impact of programming language features on LLM-generated code quality, highlighting how language-specific attributes influence code reliability. 
%\MARIO{could we add a brief description of the main results?}

While the aforementioned pieces of work have assessed LLM-generated code quality from various perspectives, they overlooked the influence of multilingual prompts. This seemingly simple factor can have profound implications for code quality and equitable access to AI-powered development tools, making it essential to investigate given the global reach of software development and the multilingual nature of LLM training data.
%extent to which the quality is influenced by prompts written in different  languages.

\subsection{Multilingual Inconsistencies in LLMs}
\label{sec:multi}
%LLMs have been extensively studied for their ability to generate code at varying levels of abstraction, ranging from high-level source code to low-level programming constructs \cite{li2022unleashing,cummins2023large}. Also, several studies have explored the efficacy of LLMs in generating code across diverse programming languages, spanning both high-level and low-level paradigms \cite{chai2024mceval, yan2023codescope}. 
Understanding the adaptability of LLMs across linguistic contexts remains underexplored, from high-resource languages such as English to other forms of languages, including those with a different alphabet and/or writing systems (\eg, Chinese, Hindi). If left unexamined, this disparity could reflect the performance gaps observed across various tasks, such as logical reasoning \cite{foroutan2023breaking}, natural language understanding \cite{huang2023not}, and natural language generation \cite{deng2023multilingual}.
Shi \etal \cite{shi2022language} highlighted these discrepancies by demonstrating that models such as GPT-3 \cite{brown2020language} and PaLM \cite{anil2023palm} exhibited significantly lower accuracy when answering mathematical questions in underrepresented languages than when responding to the same queries in English. This lower accuracy can be due to the linguistic imbalance in training and alignment data, mostly in English \cite{alves2024tower,bang2023multitask}. As we show in our study, possible consequences may occur in software engineering applications as well.
%where models such as GPT-3 \cite{brown2020language} are predominantly trained on English-language corpora \cite{alves2024tower,bang2023multitask}. 
%This imbalance has implications that go beyond general NLP tasks, raising concerns about the performance of LLMs when used by developers who primarily interact with them in languages other than English.
Koyanagi \etal~\cite{koynagi24} performed a preliminary study considering (i) a Japanese programming context translated into Chinese and English, and (ii) GitHub Copilot to generate code elements. They evaluated the correctness of the produced code and found that Japanese was on par with English and even outperformed it, while Chinese did not perform well.
Our work goes beyond such an analysis, as we consider multiple programming languages, four languages, and three LLMs. Also, we look at code quality from other perspectives---\ie metrics, static analysis warnings, and lexicon---beyond its correctness.

Moreover, we have avoided using similar benchmarks to avoid possible biases/leakages (\eg a Japanese programming contest may bias results towards Japanese). Instead, we leverage a benchmark based on real-world source code tasks, for which, especially for Python, Chinese turned out to perform well.
Peng \etal~\cite{humanevalxl2024peng} automatically translated the HumanEval benchmark~\cite{humaneval} into twenty-three natural languages for prompts and twelve programming languages for code. They then evaluated performance using the pass@1 metric across large language models (LLMs) of varying sizes. 
Our work introduces several novel contributions in comparison. First, we adopt CoderEval, a recent and more challenging benchmark that is currently one of the primary references for code generation, whereas Peng \etal rely on HumanEval, a benchmark that has been largely outperformed by contemporary models~\cite{paperswithcode}. Second, our evaluation goes beyond functional correctness by incorporating additional dimensions of code quality, including metric-based analysis, static analysis warnings, and lexical diversity. Third, we consider state-of-the-art LLMs that largely outperform those used by Peng \etal (they used GPT 3.5-turbo, and other models of 13B parameters or less).
Last, and more importantly, we manually validated each prompt translation generated by GPT-4, ensuring accuracy and preserving intent. In contrast, the translations in~\cite{humanevalxl2024peng}, both for prompts and code, were not manually verified, potentially introducing significant bias.
A possible approach to deal with prompt language bias is fine-tuning. However, this has been applied only on a relatively small model (T5-base~\cite{2020t5}, having 220M parameters) \cite{ernie2023chai} and may not be feasible on state-of-the-art models. Also, such a fine-tuning may create overfitting issues \cite{ernie2023chai}.

Compared to previous literature, our paper aims to investigate this issue through a systematic analysis that identifies potential biases and examines the quality of LLM-generated code in multilingual contexts. This analysis will contribute to enhancing fairness and accessibility in AI-driven SE tools, ensuring that developers across different linguistic backgrounds can effectively utilize these technologies, regardless of their primary language.
\section{Study Design} 
\label{sec:design}
The \emph{goal} of our study is to investigate the quality of LLM-generated source code when developers write queries using different languages. The \emph{perspective} is of developers who want to leverage (natural language) specifications written in their own language to generate source code with LLMs.  The \emph{context} consists of (i) 460 (230 for Python and 230 for Java) development tasks from the CoderEval~\cite{hao2024codereval} benchmark, translated into four languages:
%source code gene
%Our study investigates how prompts written in different languages affect the code quality generated by large language models (LLMs), specifically GPT-4o-mini, Claude, and DeepSeek. We compare code quality metrics regarding correctness (i.e., whether the code passes test cases), complexity, size, code smells, and static code violations. In addition to English, we analyze semantic-equivalent prompts in four other languages: 
Chinese, Hindi, Spanish, and Italian; and (ii) three state-of-the-art LLMs, \ie \gpt,  \ds, and \cl.

%This approach enables us to examine the impact of linguistic diversity on code generation and reveal potential biases embedded in the models. Understanding how different languages influence code generation is crucial for evaluating the robustness and fairness of LLMs in multilingual software engineering environments.

The study addresses three research questions:

\newcommand{\rqone}{How does a prompt written in a language other than English impact code generation in terms of correctness?}
\newcommand{\rqtwo}{How does a prompt written in a language other than English affect code quality metrics, design issues, and static code analysis tool violations?}
\newcommand{\rqthree}{How does a prompt written in a language other than English impact the source code lexicon?}

\textbf{RQ$_1$}: \emph{\rqone} 
We assess whether task specifications (and consequently prompts) written in different languages affect the functional correctness of the generated code, hence possibly requiring better testing and code review.
%Comparing results across languages provides insights into how LLMs interpret functional requirements differently.

\textbf{RQ$_2$}: \emph{\rqtwo} 
We (i) look at whether the code is structurally different and more or less commented by examining source code metrics, and (ii) check code quality aspects by relying on static analysis tools that developers typically use in their daily work. 

%This question examines the impact of linguistic differences on various aspects of code quality, including code complexity, size (e.g., lines of code, lines of comments), design flaws, and static code issues (e.g., code smells, unused imports, resource leaks). Using established static analysis tools, we aim to identify patterns in how different languages influence the structure, efficiency, and integrity of the generated code, offering a comprehensive view of LLMs' performance across linguistic contexts.

\textbf{RQ$_3$}: \emph{\rqthree} 
We look at the language of identifiers, comments, and string literals, as this might impact developers' further interventions to ensure proper documentation and nationalization of the resulting code.

%This question analyzes the linguistic characteristics of the code, focusing on comments, identifiers, and literals (e.g., error messages and print statements) to determine the extent to which the original language appears in the output. This analysis sheds light on whether LLMs maintain linguistic consistency or introduce unintended biases when generating code from non-English prompts.

%By addressing these questions, our study provides a comprehensive evaluation of the strengths and limitations of LLMs in generating source code from prompts written in different languages. This research contributes to the broader discourse on the applicability of LLMs in software engineering tasks. 
%The study replication package,  including analysis scripts, generated code, associated metrics, and detailed statistical results, are available in the paper's replication package\TODO{add the repository reference}.

\subsection{Context Selection: Dataset}
As a dataset for our study, we adopt CoderEval~\cite{hao2024codereval}, a benchmark designed to evaluate code generation models across two programming languages: Python and Java. It comprises 230 Python and 230 Java problems curated from open-source repositories, focusing on practical challenges that involve external dependencies, class structures, and contextual information. The problems cover a diverse range of tasks, including algorithmic solutions, object-oriented programming, and integration with external libraries, providing a good test bed for an evaluation framework. The dataset also includes test cases to assess the correctness of the generated code.

Unlike simpler benchmarks used in previous related work \cite{humanevalxl2024peng} (\eg HumanEval \cite{humaneval}), CoderEval evaluates a model's ability to generate correct and executable code by requiring an understanding of contextual elements such as APIs, variables, and type definitions. This makes it particularly suitable for assessing the real-world applicability of code generation models. This peculiar element made the benchmark extremely effective in evaluating code generator models--as highlighted by relevant research \cite{dong2024self, shen2023pangucoder,fangwen2024clarify}.
%For this reason, we focus our study on CoderEval to provide a detailed analysis using recent, practical data.

We adopted CoderEval's recommendation to utilize both Java and Python as the programming languages for our study. Also, such languages are considered among the most widely used by developers~\cite{jetbrains,githut,pypl}.

\rev{To broaden the scope of our evaluation beyond method-level generation, we complement CoderEval with ClassEval~\cite{du2023classeval}, a benchmark for class-level Python code generation. ClassEval consists of 100 classes (410 methods in total) sourced from open-source repositories. Here, rather than generating individual methods in isolation, models must produce entire class implementations -- handling constructors, methods, and the dependencies between them -- which raises the bar in terms of both complexity and coherence. We followed the same experimental protocol for ClassEval as for CoderEval: translating the task descriptions into the four target languages with manual curation by native-speaker co-authors, generating code with the same three LLMs using identical temperature and seed settings, and applying static analysis using Lizard, \flake, and \pylint. Following the ClassEval protocol, each generated method was spliced into the ground-truth class and evaluated against its method-specific test class. For Claude, we ran the full ten-iteration protocol ($n = 1{,}000$ class--iteration pairs per language), matching the CoderEval design. For GPT and DeepSeek, pass rates are reported on a single iteration ($n = 100$ classes per language), supported by the Friedman tests in Section~3.5 that showed no significant iteration-to-iteration variability for any model on CoderEval. \pmd was excluded as it is Java-only; \sonar was omitted because in the CoderEval results \pylint and \sonar produced consistent warning-count orderings and comparable effect sizes. For RQ$_1$, we report class-level pass rates (a class passes only if all its methods pass in the same iteration); for RQ$_2$ and RQ$_3$, the analysis remains at the method level to maintain comparability with the CoderEval results.}

%The 2024 JetBrains Report indicates that Python and Java are software developers' most widely used languages~\cite{jetbrains}. 
%JetBrains conducted a survey of 23,262 developers which revealed that 57\% and 46\% use Python and Java for programming tasks. Additionally,
%Also, GitHub statistics show that Python and Java are the most used programming languages, accounting for 16.92\% of repositories for Python, while Java comprises 11.7\%~\cite{githut}, and consistent results are reported by the PYPL index~\cite{pypl}.
%Java and Python are the most popular programming languages (i.e., how often language tutorials are searched on Google). Respectively, they have a 30.27\% and 14.89\% of share dated on March 2025. Given their widespread adoption, focusing on these two languages provides a comprehensive basis for evaluating LLM performance on multilingual code generation.

\subsection{Creation of Queries in Different Languages}

To assess the impact of multilingual prompts on code generation tasks, we translated 230 Python and 230 Java problems from CoderEval into four languages (chosen for the reasons explained in \secref{sec:introduction}): Chinese, Hindi, Spanish, and Italian. 
%As outlined in \secref{sec:introduction}, we selected Chinese because it is one of the most widely spoken languages globally and utilizes a logographic writing system. Hindi was chosen for its distinct Devanagari script and its prominence as one of the most spoken languages in South Asia. Lastly, Spanish and Italian were included as representatives of Latin-based languages, allowing us to examine the impact of linguistic variations within the same language family. % influence code generation.
To perform the translation, some authors, native speakers of each language, produced the queries for the 460 coding problems, being supported by an initial translation made by the GPT-4 API. However, note that this was only initial support because each query was manually refined, performing various fixes, including (i) adding quotes to the words belonging to English to maintain the original objective of the prompt, (ii) fixing mistranslation of terms and words in the prompts, and (iii) removing unwanted code or context added by GPT, within other examples. In general, the manual refinement aimed to ensure that the final set of queries represents a result that is close to human-generated queries. GPT-4 often translates English terms that have no clear equivalent in the target language, which can make the result less natural (\eg terms like ``Decorator'' were erroneously translated to ``Decoratore'' in Italian, or in Spanish ``key'' was translated in ``clave'' since it refers to a dictionary key). In some cases--such as class names--retaining the original English term would be more appropriate.
\rev{This two-stage process, combining automated translation with expert manual curation, provides an inherent form of cross-validation, as the human translator independently assessed and corrected the machine-generated output. Importantly, the translators were domain experts and co-authors with active research experience in software engineering, which significantly reduces the risk of misinterpreting technical jargon or programming-specific terminology.}

%\MAX{simply describe here the kinds of fixes being performed, making clear that in the end the resulting query was as a human would have done it}

\subsection{Code Generation through LLMs}

The starting phase of our study focuses on generating the source code for the selected programming languages (Java and Python). We have selected three well-established LLMs to generate the code: \rev{\gpt}, DeepSeek-V3, and Claude-3.5.

%\begin{compactenum}
\rev{\textbf{\gpt}} is a state-of-the-art LLM developed by OpenAI, representing a significant advancement over its predecessor, GPT-3.5~\cite{gpt4}. 
Notably, it exhibits enhanced capabilities in handling multi-modal data, including code, which translates into improved performance across a broader range of tasks, such as software engineering-related practices. 
Recent literature has shown its effectiveness in various software engineering applications~\cite{xueying2024evaluating, li2025structured, zheng2024open, cipriano2024llms, serafini2024chatgpt, midolo2025automated}. The specific model used in this study is gpt-4o-mini-2024-07-18.

 \textbf{DeepSeek}~\cite{deepseekai2025report} is an LLM developed with an emphasis on code understanding and generation. 
%For software development tasks, it excels at complex tasks such as code translation, bug detection, and performance optimization using advanced transformer-based architectures. 
Recent studies underscore its effectiveness in both code generation and the analysis of large codebases\cite{deepseekai2025report, deepseek2024scaling, deepseek2025incentivizing}. The specific model used in this study for code generation is \textit{deepseek-chat (DeepSeek-V3)}, accessed via APIs.

\textbf{Claude} is a state-of-the-art large language model developed by Anthropic~\cite{AnthropicClaude3}. 
%It is known for its ability to assist with various tasks, including code generation. 
Claude has shown strong performance in generating functional code, debugging, and refactoring across different programming languages~\cite{sobo2025evaluating, nath2024wip, murr2023testingllmscodegeneration, jiang2024surveylargelanguagemodels}. The specific model used in this study is \textit{Claude 3.5} (\ie claude-3-5-sonnet-20241022).
%\end{compactenum}

\textbf{Prompt Engineering}. The prompt for an LLM can be divided into two components: the \texttt{system prompt} and the 	\texttt{user prompt}. The system prompt defines the model's behavior through specific instructions, while the user prompt contains the actual input requesting a task or response. The formulation of these prompts significantly impacts the LLM's code generation performance~\cite{liu2023improving, li2024acecoder, li2025structured}. Following best practices outlined in prior work~\cite{shinn2023advances}, we designed our system prompt as shown in the next box. This prompt includes two customizable tags: ``prog\_lang'' specifies the target programming language, and ``self\_contained\_class'' instructs the model to generate self-contained classes--particularly for Java--to ensure the output is compilable. This requirement is essential for utilizing static analysis tools such as \sonar.

%\ANTONIO{We can drop right here an example where we show the curation process}

For \gpt and \ds, we use the instruction:  ``\textit{The code must be self-contained, including imports and dependencies.}'' For Claude, we adapted this to: ``\textit{Please generate the method within a class so that it can be compiled, including imports and dependencies},`` as the original phrasing did not consistently produce class-level outputs. For the \textit{user prompt}, we adopt the methodology proposed in the CoderEval dataset~\cite{hao2024codereval}, using the method signature and corresponding docstring as input for code generation. \rev{No retrieval mechanisms or supplementary repository-level context (\eg cross-file dependencies, surrounding class definitions, or existing comments from the source repository) were incorporated during the generation process. This design choice ensures a controlled comparison across prompt languages, avoiding the introduction of English-language bias from repository-level artifacts that are predominantly written in English.}

% \begin{promptbox} 
% \textbf{System}: ``You are an AI that only responds with [prog\_lang] code. You will be given a function signature and its docstring by the user. Write your full implementation (restate the function signature).
% [self\_contained\_class].
% Use a [prog\_lang] code block to write your response. For example:
% ```[prog\_lang]
% print(``Hello World!``)
% `````
% \end{promptbox}

\begin{tcolorbox}[colback=blue!5!white, colframe=blue!50!black, 
  title=\textbf{System Prompt Used for Generation}]
\ttfamily\textbf{System}: \textit{You are an AI that only responds with \texttt{[prog\_lang]} code. 
You will be given a function signature and its docstring by the user. 
Write your full implementation (restate the function signature)}. 
\texttt{[self\_contained\_class]}.

\textit{Use a \texttt{[prog\_lang]} code block to write your response. 
For example:}

\begin{lstlisting}[
    language=Python,
    backgroundcolor=\color{gray!5},
    basicstyle=\ttfamily\mdseries,
    keywordstyle=\color{blue}\bfseries,
    stringstyle=\color{red},
    commentstyle=\color{codegreen}\itshape,
    showstringspaces=false,
    frame=single,
    rulecolor=\color{gray!40},
    xleftmargin=8pt,
    xrightmargin=8pt
]
[prog_lang]
print("Hello  World!")
\end{lstlisting}

\end{tcolorbox}

\textbf{Generating the Code}. In line with recent work~\cite{liu2024refining, dong2024self, xueying2024evaluating}, we conducted the generation process with temperature zero and a fixed seed. \rev{The specific API parameters used for each model are: (i)~GPT-4o mini (\texttt{gpt-4o-mini-2024-07-18}): temperature = 0, seed = 2025; (ii)~DeepSeek-V3 (\texttt{deepseek-chat}): temperature = 0, seed = 2025; (iii)~Claude 3.5 Sonnet (\texttt{claude-3-5-sonnet-20241022}): temperature = 0. At the time of our experiments, the Anthropic API did not support a user-specified seed parameter; non-determinism was instead mitigated through ten repeated iterations per task
Prior work has demonstrated that lower temperatures yield higher functional correctness in code generation~\cite{arora2024optimizing} and produce more consistent outputs~\cite{doderlein2023piloting}.} This is because low temperature settings reduce randomness in model outputs compared to default configurations, leading to more consistent and reproducible results~\cite{ouyang2025empirical}. 
Also, research has shown that a low temperature improves performance in code generation tasks~\cite{arora2024optimizing, doderlein2023piloting}. Since these models do not ensure determinism even with a low temperature, we executed ten iterations for each generation task, \rev{yielding a total of 69,000 generated code samples (460~tasks $\times$ 5~languages $\times$ 3~LLMs $\times$ 10~iterations). This design explicitly accounts for the inherent non-determinism of LLMs, even at temperature zero~\cite{ouyang2025empirical},} and kept, for the analysis, the output of each iteration. 
%\ANTONIO{Did we select the value occurring the most? If yes, this can be introduced as a self-consistency prompting.}

\subsection{Code Quality Analysis}\label{sec:codequality}
Our study analyzes the impact of prompt language on code generation quality through two key perspectives: quantitative code metrics (cyclomatic complexity, lines of code, comment ratios) and qualitative defect detection (code smells, resource leaks). We employ standard static analysis tools to systematically compare how different languages influence the structural and functional characteristics of LLM-generated code.

\textbf{Choice of static analysis tools and metric extractors}. 
\rev{Our selection of code quality metrics and static analysis tools is guided by established software quality frameworks and survey literature. The ISO/IEC 25010 standard~\cite{iso25010} defines software product quality along characteristics including maintainability, reliability, and security, all of which are assessed by our selected tools and metrics. Nunez-Varela~\etal~\cite{nunez2017source} provide a comprehensive mapping of source code metrics, identifying complexity, size, coupling, and cohesion as the primary measurement dimensions. Our metric selection, comprising cyclomatic complexity (CC), cognitive complexity (Cogn), and non-comment lines of code (NLOC), covers the two most widely adopted dimensions (complexity and size) from their taxonomy.}

\rev{For static analysis tool selection, we adopted the following criteria: (i)~support for at least one of the target programming languages (Python or Java), (ii)~configuration independence, to ensure that results are not influenced by project-specific rule sets and to allow fair cross-language comparison, (iii)~complementary coverage of different quality dimensions, including complexity, coding style, defects, and security, and (iv)~established validation in academic literature, ensuring trustworthiness and recognized effectiveness in code quality assessment research~\cite{siddiq2023generate, yetiştiren2023evaluatingcodequalityaiassisted}. Based on these criteria,}
we rely on \pylint~\cite{pylint} and \flake~\cite{flake8} for Python, \pmd~\cite{copeland2005pmd} for Java, \sonar~\cite{sonarcloud} and \textsc{lizard}~\cite{lizard} for both languages. 

\begin{itemize}
    \item \pylint~\cite{pylint} is a popular and comprehensive static analysis tool that enforces coding standards and detects various issues in Python code~\cite{siddiq2022empirical}. 
\textsc{flake8} is another widely used tool for Python, which checks for syntax errors, style issues, and code complexity. 
%\BOWEN{it's unclear what does ``these tools'' in the following sentence referring to?}
    Together, Pylint and Flake8 enable us to assess Python code quality from multiple dimensions, beyond just functional correctness~\cite{liu2024refining}.
    \item \pmd~\cite{copeland2005pmd} is a well-known static analysis tool that inspects Java source code to identify potential problems and provides suggestions for improvements~\cite{liu2024refining, wattanakriengkrai2022predicting}. 
%This tool evaluates the Java source code against a set of rules, reporting warnings for violations, their priority, and the corresponding lines in the file.
    \item \sonar~\cite{sonarcloud} is a cloud-based static analysis tool that helps developers identify and fix code quality issues in multiple programming languages. It detects bugs, code smells, security vulnerabilities, and duplications, providing insights into code maintainability and reliability~\cite{puspaningrum2023vulnerable, nocera2024training}.
    \item \textsc{lizard}~\cite{lizard} is a lightweight and fast static analysis tool that extracts code metrics
%measures code complexity \MAX{can't we say it's a metric extraction} 
across multiple programming languages, including Python and Java. 
%It helps identify complex functions that may be harder to maintain or test, making it a valuable tool for assessing code structure and improving software quality.
\end{itemize}

The choice of these static analysis tools is motivated by their ability to provide diverse perspectives on code quality across both languages, hence enhancing the reliability of our analyses. Additionally, these tools are well-established and validated in academic literature, signaling their trustworthiness and established effectiveness in code quality assessment research \cite{siddiq2023generate, yetiştiren2023evaluatingcodequalityaiassisted}. %\ANTONIO{Saima, add here}

Before moving forward, \rev{we excluded CheckStyle~\cite{checkstyle} from our analysis because its output is highly dependent on user-defined configuration rules, violating criterion~(ii) above. Including CheckStyle would have introduced variability attributable to configuration choices rather than to the prompt language effects we aim to measure. This decision is consistent with recent empirical studies on LLM-generated code quality, where CheckStyle was similarly excluded due to its heavy dependence on team-specific style configurations, which could introduce unwanted bias into the interpretation of results~\cite{afrin2025quantization}.}

\begin{table}[ht]
    \centering
    \caption{Quality indicators from static analysis tools%\BOWEN{smells and bugs by sonarcloud could be warnings in PMD, PYLINT, FLAKE8. can we distinguish error and warning in table 2 and here, we breakdown  warnings in each of PMD, PYLINT, FLAKE8 into smell and warnings of PMD, smell and warnings of PYLINT, smell and warnings of FLAKE8?}\ALESSANDRO{FIXED}
    }
    \label{tab:metrics}
    \begin{tabular}{ll}
    \toprule
    \textbf{Metric/Indicator} & \textbf{Description} \\
    \midrule
    \nloc (Lizard) & Non-Comment Lines of Code \\
    \cc (Lizard) & Cyclomatic Complexity \\
    \cmt (SonarCloud) & \# of Comment Lines \\
    \cmd (SonarCloud) & Comment Density \\
    \cogn (SonarCloud) & Cognitive Complexity (Code Understandability) \\
    \smells (SonarCloud) & Code Design Flaws \\
    \bugs (SonarCloud) & Code Defects and Errors \\
    \smells (\pmd) & \rev{Style, Design, and Maintainability Issues}\\
    \warnings (\pmd) & \rev{Bug-Prone and Concurrency Issues}\\
    \smells (\pylint) & \rev{Convention and Refactoring Issues}\\
    \warnings (\pylint) & \rev{Errors, Faults, and Bug-Prone Code}\\
    \smells (\flake) & \rev{Style and Documentation Issues}\\
    \warnings (\flake) & \rev{Bug-Prone Patterns and Runtime Faults}\\
    \bottomrule
    \end{tabular}
    % \vspace{-2mm}\\
    
\end{table}

Table \ref{tab:metrics} summarizes the indicators provided by the different tools. Note that for comment lines, we decided to count both the absolute number and the density.
For what concerns \pmd, \pylint, \flake, and \sonar warnings, in RQ$_2$ we first count the absolute number of warnings contained in each generated code function, and then we perform an in-depth analysis on warning types. 

Additional clarification is required for the \sonar metrics. Cognitive Complexity measures code understandability, aligning with developer intuition to identify challenging modules to maintain~\cite{campbell2018cognitive}. Code smells refer to design flaws or implementation patterns that, while not causing immediate failures, increase maintenance difficulty. Examples include code duplication, long methods, complex conditionals, and inconsistent naming. Bugs denote code defects likely to cause incorrect behavior or runtime errors, such as null pointer dereferences, logical flaws, and off-by-one errors~\cite{lenarduzzi2020sonarqube, nocera2025software}.

\begin{table}[ht]
\centering
\caption{Smells and Warnings detected by static analysis tools}
\label{tab:errordescription}
\resizebox{\textwidth}{!}{%
\begin{tabular}{@{} l l l l l @{}}
\toprule
\textbf{Tool} & \rev{\textbf{Severity}} & \textbf{Category} & \textbf{Name} & \textbf{Description (single example)} \\
\midrule
\flake & \rev{Style} & Smell  & PycodestyleWarnings    & Minor PEP 8 issue, e.g., overly long line. \\
\flake & \rev{Style} & Smell  & PycodestyleErrors      & PEP 8 error, e.g., bad indentation. \\
\flake & \rev{Style} & Smell  & DocstringIssues        & Missing or incomplete docstring, e.g., no return description. \\
\flake & \rev{Warning} & Smell  & McCabeComplexity       & Function too complex, e.g., many nested \texttt{if} statements. \\
\flake & \rev{Warning} & Warning & BugRisks.BestPractices & Bug-prone pattern, e.g., mutable default argument. \\
\flake & \rev{Error} & Warning & PyFlakesErrors         & Likely runtime fault, e.g., undefined variable. \\
\midrule
\pylint & \rev{Fatal} & Warning & fatal                  & Stops analysis, e.g., syntax error. \\
\pylint & \rev{Error} & Warning & error                  & Likely runtime failure, e.g., wrong argument count. \\
\pylint & \rev{Warning} & Warning & warning                & Bug-prone code, e.g., broad \texttt{except} clause. \\
\pylint & \rev{Convention} & Smell  & convention             & Breaks style conventions, e.g., class name not in PascalCase. \\
\pylint & \rev{Refactor} & Smell  & refactor               & Code can be simplified, e.g., duplicate logic. \\
\midrule
\sonar & \rev{Blocker} & Warning & BUG.BLOCKER            & Certain crash, e.g., unhandled fatal exception. \\
\sonar & \rev{Critical} & Warning & BUG.CRITICAL           & High-impact bug, e.g., resource leak. \\
\sonar & \rev{Major} & Warning & BUG.MAJOR              & Wrong result, e.g., off-by-one error. \\
\sonar & \rev{Minor} & Warning & BUG.MINOR              & Low-severity defect, e.g., redundant condition. \\
\sonar & \rev{Blocker} & Smell  & CODE.SMELL.BLOCKER     & Blocks maintenance, e.g., 500-line duplicated method. \\
\sonar & \rev{Critical} & Smell  & CODE.SMELL.CRITICAL    & Hard to maintain, e.g., very deep nesting. \\
\sonar & \rev{Major} & Smell  & CODE.SMELL.MAJOR       & Significant maintainability issue, e.g., long parameter list. \\
\sonar & \rev{Minor} & Smell  & CODE.SMELL.MINOR       & Minor maintainability concern, e.g., inconsistent naming. \\
\sonar & \rev{Info} & Smell  & CODE.SMELL.INFO        & Informational note, e.g., optional style suggestion. \\
\sonar & \rev{Critical} & Warning & SecurityHotspot        & Possible security risk, e.g., hard-coded password. \\
\midrule
\pmd   & \rev{P4--Low} & Smell  & codestyle              & Style violation, e.g., inconsistent brace placement. \\
\pmd   & \rev{P4--Low} & Smell  & documentation          & Missing or poor Javadoc, e.g., no @param tags. \\
\pmd   & \rev{P3--Medium} & Smell  & best.practices         & Violates Java best practices, e.g., unclosed stream. \\
\pmd   & \rev{P3--Medium} & Smell  & design                 & Single-function design smell, e.g., too many parameters. \\
\pmd   & \rev{P4--Low} & Smell  & performance            & Inefficient code, e.g., using \texttt{String +} in loops. \\
\pmd   & \rev{P1--High} & Warning & error.prone            & Bug-prone code, e.g., \texttt{==} instead of \texttt{.equals}. \\
\pmd   & \rev{P1--High} & Warning & multithreading         & Concurrency hazard, e.g., double-checked locking; N/A if no threading. \\
\bottomrule
\end{tabular}%
}
\vspace{-2mm}
\end{table}

Table \ref{tab:errordescription} summarizes the errors and warnings provided by the different static analysis tools. We have grouped the errors according to the categories provided by the documentation of each tool. \rev{Note that in Table~\ref{tab:errordescription}, we adopt a unified two-category classification across all tools: ``Warning'' denotes potential defects, bug-prone patterns, and runtime faults, while ``Smell'' denotes style, maintainability, and documentation concerns. Accordingly, SonarCloud entries categorized as ``Warning'' in Table~\ref{tab:errordescription} (BUG.BLOCKER through BUG.MINOR and SecurityHotspot) correspond to the ``Bugs'' indicator reported in Table~\ref{tab:metrics}.} Such fine-grained classification aims to shed light on the differences in terms of code quality for each language as well as the magnitude of such differences (if any).

\textbf{Analysis Pipeline}. The generated code undergoes an automated pipeline that extracts all the metrics required for analysis. 
First, to address RQ$_1$, we computed the pass@k metric for each query and for each of the 10 LLM-generated solutions.
%whether all test cases passed or not\BOWEN{pass@K is the common name of the metric}. 
%\emph{Pass@k} metric by testing all generated entries. This metric evaluates whether LCMs \MARIO{We have not explained in the paper what LCMs are} pass all unit tests within $k$ solutions. As previously discussed, we set $k$ to ten, generating ten versions of the same problem to mitigate non-determinism issues. 
We execute all tests within the Docker environment provided by the CoderEval benchmark~\cite{hao2024codereval}.
Second, we applied static analysis tools to extract quality metrics and warnings. We executed \textsc{lizard}, \pylint, and \flake through their command-line interfaces.  We ran \pmd and \sonar using GitHub Actions. %environment, which provides an automated solution for running static analysis tools on newly committed code. 
Note that, for Java, \sonar works on bytecode. Therefore, the code must be compiled. Note that all cases for which the code does not compile are skipped from this analysis.
\rev{To provide a complete picture of code quality, Table~\ref{tab:tests} also reports the Java compilation success rates across all prompt languages and LLMs. For GPT and DeepSeek, compilation rates remain generally consistent across prompt languages, ranging from 50.57\% to 54.74\% and from 44.26\% to 48.22\%, respectively, suggesting that prompt language does not introduce a systematic compilation bias for these models. Claude, however, exhibits a notable exception: Hindi prompts yield a compilation rate of only 33.91\%, substantially lower than the 46.61\%--50.87\% observed for the other languages. This result is consistent with the significantly lower functional correctness reported for Claude-Hindi in RQ$_1$. Accordingly, the static analysis results for Claude-Hindi in RQ$_2$ should be interpreted with caution, as they are derived from a reduced subset of compilable samples. Compilation failure itself constitutes an additional quality indicator, as it reflects syntactic or structural issues in the generated code. Compilation rates are reported only for Java, since Python is an interpreted language and does not require a separate compilation step.}

\subsection{Manual Analysis of Identifiers, Comments, and Literals}
\label{sec:rq3method}
To address RQ$_3$, we manually inspect the generated code to identify the language used for (i) identifiers, (ii) comments, and (iii) string literals. To make such an analysis doable, we performed it on one (first) iteration only of each LLM. 
To determine whether analyzing a single iteration would constitute a serious threat, we studied the variability of indicators (metrics and static analysis warnings) computed for RQ$_2$ using Friedman's test \cite{friedman1937}, which is a test for dependent variables, allowing us to separate the within-subject effect (iteration) from the between-subject effect (metric variability across different tasks). The test \rev{never revealed statistically significant differences across iterations for any of the computed indicators ($p > 0.05$ in all cases), confirming that a single iteration provides a representative sample for the manual lexicon analysis. Full test statistics are available in the replication package.}

One of the authors (a native speaker of the respective language) inspected each generated code element and classified it along the dimensions above. It is important to note that, for the identifiers (especially function name/parameters), the model is likely to be biased by the benchmark specification, \ie the signature contained in the benchmark. Such a signature could not be translated, as this would have impacted the usage of test cases.  In some circumstances, string literals can also be from the specification.

Given the relatively simple task yet the relatively large number of samples to inspect for each language \rev{($230 \cdot 2 \cdot 3 = 1{,}380$ for CoderEval and $410 \cdot 3 = 1{,}230$ for ClassEval)}, we only assess the reliability of the coding task by performing a second (independent) manual validation on a sample of 50 randomly generated samples for each programming language, and computed Cohen's $k$ \cite{kappa} inter-rater agreement. The agreement was almost perfect for identifiers, over 0.77 for comments, and over 0.64 for literals, very strong in any case (see details in the replication package). Therefore, we could proceed with one evaluator only.

It is important to point out that the analysis in RQ$_3$ must be considered with particular care for two reasons:
\begin{enumerate}
\item The benchmark prompts contain signatures in English. This is necessary to make test executables and to preserve dependencies with the rest of the code in which the generated methods will be integrated. 
\item While one may expect comments and literals in any of the considered languages, for Hindi and Chinese, the translation of identifiers does not make sense.
\end{enumerate}

To mitigate this limitation,  we conduct a controlled experiment 
using 100 randomly selected Python examples translated into Italian, assessing the extent to which modifying the function signature impacts consistency in identifier and comment generation.

We then run the query against the LLMs with two different configurations. Firstly, we translate the signature/parameter specification into Italian. We have chosen Italian because translating function signatures in a language that includes scripts (\eg Chinese), would create a mismatch in the reference alphabet. Secondly, we keep the signature in English, but we clearly ask in the system prompt to use comments and identifiers in Italian, \ie we added this sentence in the system prompt: \emph{``Comments and identifiers must be in Italian.''}

\subsection{Analysis Methodology}
In the following, we describe the statistical analyses conducted to address the different RQs. Such analyses were conducted using the R \cite{R} statistical environment, and for all tests, we assumed a significance level $\alpha=0.05$.

To address RQ$_1$, since we have for each query and each iteration a test outcome (pass or fail), we compare the test outcome of English with the other languages using McNemar's test \cite{mcnemar}, a statistical test suitable for dependent categorical data. Due to multiple comparisons, $p$-values are adjusted using the Benjamini-Hochberg \cite{yoav:jstor1995} correction. We complemented the test with the Cohen'$g$ \cite{cohen1988} effect size measure, a non-parametric effect size for paired categorical data. \rev{The classification of $g$ values into effect size categories is reported in Table~4.}

% \begin{compactenum}
% \item $0.00 \le g< 0.05$: (E)xtra (S)mall;
% \item $0.05 < 0.15$: (S)mall;
% \item $0.15 < 0.25$: (M)edium.
% \item $\ge0.25$: (L)arge.
% \end{compactenum}
%), which is the one suitable for that test. 

\begin{table*}[t!]
\centering
\footnotesize
\caption{\rev{Percentages of passed tests across languages and benchmarks.} $g$ refers to the Cohen's $g$ \cite{cohen1988} effect size. Values in boldface and with a ``*'' indicate that McNemar's test found a statistically significant difference. \rev{For Java, we also report the compilation success rate (Comp. \%), computed over 2,300 samples (230 tasks $\times$ 10 iterations). Compilation rates are not reported for Python, as it is an interpreted language and does not require a separate compilation step.} \rev{For ClassEval, we report class-level pass rates: a class passes only if all its methods pass in the same iteration. Claude pass rates aggregate across the ten iterations (1{,}000 class--iteration pairs per language), while GPT and DeepSeek pass rates are reported on the single iteration (100 classes per language).}}
\label{tab:tests}
\resizebox{\textwidth}{!}{%
\begin{tabular}{l l r >{\centering\arraybackslash}p{1cm} | r >{\centering\arraybackslash}p{1cm} r | r >{\centering\arraybackslash}p{1cm} r | r >{\centering\arraybackslash}p{1cm} r | r >{\centering\arraybackslash}p{1cm} r}
\toprule
& & \textbf{English} & & \multicolumn{3}{c|}{\textbf{Chinese}} 
& \multicolumn{3}{c|}{\textbf{Hindi}} 
& \multicolumn{3}{c|}{\textbf{Spanish}} 
& \multicolumn{3}{c}{\textbf{Italian}} \\
& \textbf{Model} & \textbf{\% pass} & \rev{\textbf{Comp.}}
& \textbf{\% pass} & \rev{\textbf{Comp.}} & \textbf{g} 
& \textbf{\% pass} & \rev{\textbf{Comp.}} & \textbf{g} 
& \textbf{\% pass} & \rev{\textbf{Comp.}} & \textbf{g} 
& \textbf{\% pass} & \rev{\textbf{Comp.}} & \textbf{g} \\
\midrule
% PYTHON
& \textbf{\gp} & 23.35 & \rev{--} & 30.00 & \rev{--} & \textbf{0.28 (L) *} & 22.35 & \rev{--} & 0.08 (S) & 24.09 & \rev{--} & 0.07 (S) & 23.91 & \rev{--} & 0.04 (XS) \\
\textbf{Python} & \textbf{\ds} & 26.87 & \rev{--} & 30.00 & \rev{--} & \textbf{0.19 (M) *} & 25.96 & \rev{--} & \textbf{0.10 (S) *} & 24.13 & \rev{--} & \textbf{0.37 (L) *} & 25.30 & \rev{--} & \textbf{0.20 (M) *} \\
& \textbf{\cl} & 25.13 & \rev{--} & 31.04 & \rev{--} & \textbf{0.25 (L) *} & 25.26 & \rev{--} & 0.01 (XS) & 24.48 & \rev{--} & 0.05 (S) & 25.17 & \rev{--} & $<0.01$ (XS) \\
\midrule
% JAVA
& \textbf{\gp} & 32.78 & \rev{51.52} & 31.2 & \rev{54.13} & \textbf{0.12 (S) *} & 33.61 & \rev{54.74} & 0.08 (S) & 32.26 & \rev{50.57} & 0.03 (XS) & 33.39 & \rev{52.61} & 0.05 (S) \\
\textbf{Java} & \textbf{\ds} & 26.35 & \rev{45.43} & 26.83 & \rev{45.35} & 0.03 (XS) & 25.70 & \rev{44.57} & 0.06 (S) & 26.83 & \rev{44.26} & 0.03 (XS) & 28.43 & \rev{48.22} & \textbf{0.21 (M) *} \\
& \textbf{\cl} & 37.13 & \rev{49.04} & 32.30 & \rev{46.61} & \textbf{0.27 (L) *} & 17.91 & \rev{33.91} & \textbf{0.44 (L) *} & 35.00 & \rev{47.13} & \textbf{0.09 (S) *} & 37.26 & \rev{50.87} & $<0.01$ (XS) \\
\midrule
% CLASSEVAL (NEW — class-level)
& \rev{\textbf{\gp}} & \rev{37.00} & \rev{--} & \rev{36.00} & \rev{--} & \rev{0.06 (S)} & \rev{38.00} & \rev{--} & \rev{0.06 (S)} & \rev{34.00} & \rev{--} & \rev{0.14 (S)} & \rev{33.00} & \rev{--} & \rev{0.20 (M)} \\
\rev{\textbf{ClassEval}} & \rev{\textbf{\ds}} & \rev{41.00} & \rev{--} & \rev{45.00} & \rev{--} & \rev{0.14 (S)} & \rev{39.00} & \rev{--} & \rev{0.08 (S)} & \rev{36.00} & \rev{--} & \rev{0.19 (M)} & \rev{38.00} & \rev{--} & \rev{0.12 (S)} \\
& \rev{\textbf{\cl}} & \rev{12.80} & \rev{--} & \rev{11.30} & \rev{--} & \rev{0.16 (M)} & \rev{10.20} & \rev{--} & \rev{\textbf{0.25 (L) *}} & \rev{10.30} & \rev{--} & \rev{\textbf{0.19 (M) *}} & \rev{8.30} & \rev{--} & \rev{\textbf{0.46 (L) *}} \\
\bottomrule
\end{tabular}%
}
\vspace{-2mm}
\end{table*}

\begin{table}[h!]
\centering
\caption{Classification of $g$ values.}
\begin{tabular}{c|c}
\hline
Range of $g$ & Category \\
\hline
$0.00 \le g < 0.05$ & Extra Small \rev{(XS)} \\
$0.05 < g < 0.15$ & Small (S) \\
$0.15 < g < 0.25$ & Medium (M) \\
$g \ge 0.25$ & Large (L) \\
\hline
\end{tabular}
\end{table}

%In addition, we study the influence of the model generation runs on such an analysis using the Cochran Q test \cite{cochran1950}, which allows to analyze the within-subject (\ie multiple generation runs for the same query) effect for the same query in presence of a dependent categorical variable.

To address RQ$_2$, we have a numerical value for each generated code sample, representing either a metric or the number of warnings produced by a static analysis tool. Therefore, in this case, we perform, for each metric or static analysis tool outcome, a comparison between English and other languages using Wilcoxon signed-rank test \cite{wilcoxon} (again adjusting $p$-values), and a paired Cliff's delta ($d$) effect size \cite{Cliff:2005}. 
While we computed Cliff's $d$, it is important to remark that we do not expect tangible differences for metrics (except, possibly, for aggregated indicators such as the number of warnings from \pylint, \flake,  or \pmd.) This is because a function or a method that 
tackles the same task and is obtained from querying the model on a prompt in a different language, is unlikely to exhibit radical variations in length, complexity, or structure, yet small changes would still affect their semantics.
Therefore, we leave Cliff's $d$ for the replication package while discussing the direction of the observed differences.

To address RQ$_3$, we report and compare the number and percentage of cases for which identifiers, comments, and literals are in English, the language used for the query/prompt, or other.

% !TEX root = ../main.tex

\section{Study Results}
\label{sec:results}
%In the following, we report and discuss the results of the study.
%addressing the three research questions formulated in \secref{sec:design}.

\subsection{RQ$_1$ Linguistic Impact on Code Correctness}
\label{sub:rq1}
\tabref{tab:tests} reports statistics over the test runs for the generated code across different languages, programming languages, and LLMs. \rev{For Java, the table also includes compilation success rates, as compilation is a prerequisite for static analysis with tools such as \sonar (discussed in RQ$_2$). The pass@10 values reported in the table aggregate test outcomes across all ten iterations for each task--language--model combination, ensuring that our correctness assessment accounts for the variability inherent in LLM-based code generation.} It reports the percentage of passed tests for the four considered languages, \ie English, Chinese, Hindi, Spanish, and Italian. Moreover, Cohen's $g$ effect size is also reported with highlighted cases if McNemar's test found a statistically significant difference. $p$-values are omitted here and in the rest of the tables to ease readability, and are available in our replication package.
%The table illustrates the variation in results across all evaluated dimensions.

\noindent\textbf{Correctness in Python code generation.}
%We found that for Python code generation, all the models exhibit a significantly higher ability to handle Chinese prompts compared to English, as reflected by Cohen's $g$ values, which indicate a magnitude large for \gp, medium for \ds, and large for \cl.
We found that, \revminor{on the CoderEval Python tasks}, all the models exhibit a significantly higher ability to handle Chinese prompts compared to English, as reflected by Cohen's $g$ values, which indicate a magnitude large for \gp, medium for \ds, and large for \cl.
The performance for other languages generally aligns with that for English, except \ds, where the results are significantly worse: 26.87\% of cases passed tests for English, 25.96\% for Hindi (small effect size), 24.13\% for Spanish (large effect size), and 25.30\% for Italian (medium effect size). 

%%%% NOTE: We comment out the section below to replace with what has been asked to do in the Minor Revision

% \rev{While we cannot establish a definitive causal link without access to the models' training data, which remains undisclosed, a plausible contributing factor to the superior performance of Chinese prompts is the substantial and growing body of Chinese-language Python-related resources.} This includes a growing number of open-source repositories and technical discussions on platforms like Stack Overflow, particularly fueled by the surge in AI/ML development, where Python is dominating.

% Besides, projections for 2025 indicate that China has overtaken the United States as the global leader in software development workforce~\cite{JetBrains2025}, with an estimated 4 million developers, surpassing the U.S. by approximately 1 million. This demographic advantage suggests a broader and more diverse corpus of Chinese-language programming data, which could enhance model adaptation to Chinese-language prompts.
% Further evidence supporting China’s growing dominance in data generation comes from scientific literature \cite{liu2021rise}, which estimates that by 2025, China will produce 48.6 zettabytes of data, exceeding the total data output of the United States. Additionally, China is projected to store 27.8\% of global online data by 2025, whereas the U.S. share is expected to decline from 21\% in 2018 to 17.5\%~\cite{USNews2019}.

\revminor{Several external framings could in principle be brought to bear on this advantage: the growing body of Chinese-language Python-related resources on platforms such as Stack Overflow, the size of the Chinese software-development workforce~\cite{JetBrains2025}, and projections of Chinese-language data production~\cite{liu2021rise, USNews2019}. Each remains a plausible backdrop, though none can be verified against the training data and alignment procedures of the evaluated models, which are undisclosed.}

\begin{figure*}[htbp]
\centering
\includegraphics[width=0.9\columnwidth]{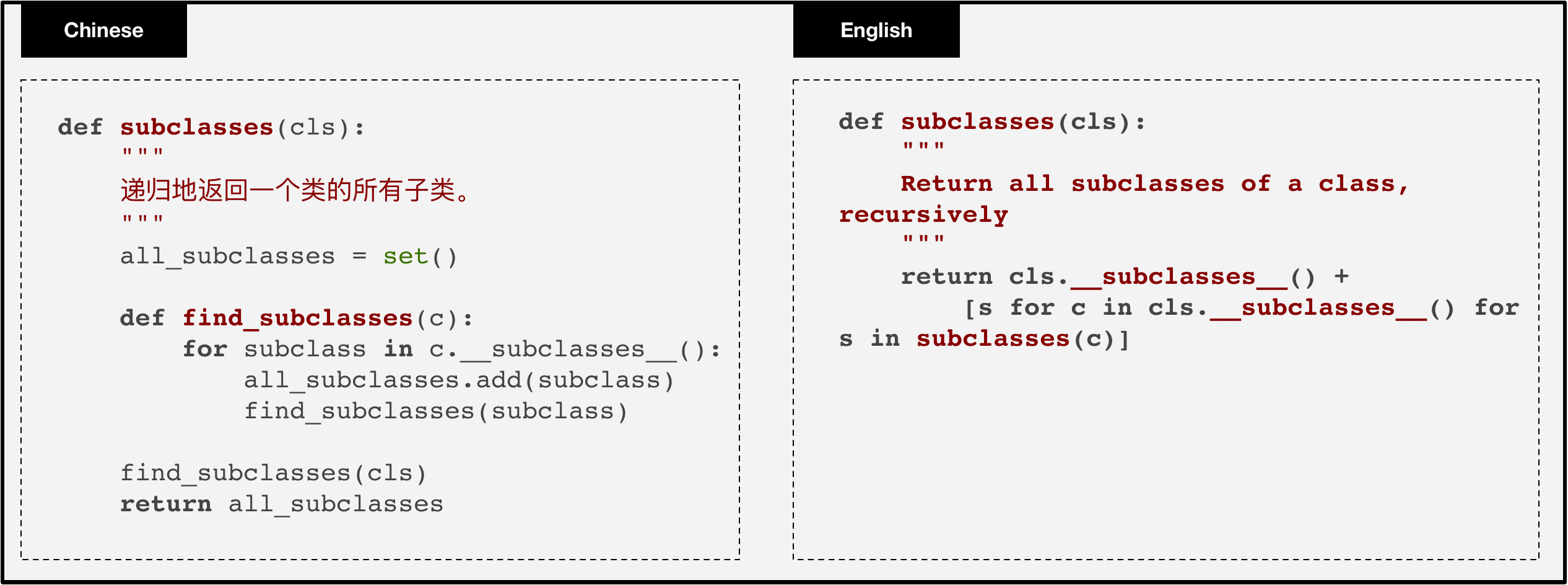}
\caption{Comparison of code generated with Chinese(left) and English(right) prompts. Only the Chinese code passes all benchmark tests.}
\label{fig:listing1}
\end{figure*}

%\ANTONIO{Saima, get rid of the listing below and replace it with pictures.} \SAIMA{done}
We provide a concrete example in Fig.~\ref{fig:listing1}. The first code is generated using a prompt in Chinese, while the second is from English. Both examples, written in Python, aim to return all subclasses of a given class. The key difference is that the Chinese-generated code uses a set to collect subclasses, which automatically removes duplicates. This approach ensures a correct handling of complex inheritance structures and avoids modifying collections during iteration. Therefore,  the Chinese-generated code successfully passed all tests, while the English-generated code failed due to two primary issues: (i) multiple calls to \texttt{cls.\_\_subclasses\_\_()}, which can lead to inconsistencies if subclasses change dynamically, and (ii) the absence of deduplication, causing duplicate subclasses in cases of shared or diamond inheritance. 
\rev{This example illustrates a recurring pattern we observed across the generated code: prompts in different languages can lead the model to select fundamentally different algorithmic strategies for the same task. In this case, the Chinese prompt led to a more defensive, set-based approach, while the English prompt yielded a more concise but less robust list-based implementation. Such differences in implementation strategy, rather than surface-level syntactic variation, are a primary driver of the correctness gaps observed in our quantitative results.}
\revminor{The same pattern recurs in the qualitative examples we develop in Section~\ref{sub:rq2}, including differences in task-requirement interpretation (Fig.~\ref{fig:bestpractices}), in style-vs.-organization trade-offs (Fig.~\ref{fig:convention}), and in the introduction of redundant control flow (Fig.~\ref{fig:bug})---strengthening the case that prompt language systematically shapes the implementation approach the model selects, rather than merely producing surface-level linguistic variation.}
%\BOWEN{can we briefly explain why it failed or which test case expose the failure?}\ALESSANDRO{Addressed}

%\vspace{-10pt}
\setlength{\textfloatsep}{10pt}

% \noindent
% \begin{minipage}[t]{0.48\textwidth}
% \begin{lstlisting}[style=mypython, caption={}, label={}]
% def subclasses(cls):
%     """
%     递归地返回一个类的所有子类。
%     """
%     all_subclasses = set()

%     def find_subclasses(c):
%         for subclass in c.__subclasses__():
%             all_subclasses.add(subclass)
%             find_subclasses(subclass)

%     find_subclasses(cls)
%     return all_subclasses
% \end{lstlisting}
% \end{minipage}%
% \hfill
% \begin{minipage}[t]{0.48\textwidth}
% \begin{lstlisting}[style=mypython, caption={}, label={}]
% def subclasses(cls):
%     """
%     Return all subclasses of a class, recursively
%     """
%     return cls.__subclasses__() +
%         [s for c in cls.__subclasses__() for s in subclasses(c)]
% \end{lstlisting}
% \end{minipage}

% \captionof{figure}{Comparison of code generated with Chinese(left) and English(right) prompts. Only the Chinese code passes all benchmark tests.}
% \label{fig:listing1}

%\begin{figure}[ht]
%  \centering
%  \includegraphics[width=0.7\linewidth]{images/Listing1.png}
%  \caption{Code generated from different language prompts: Chinese (top), and English (bottom). Only the Chinese code passes all benchmark tests.}
%  \label{lst:pythonTest}
%  \vspace{-2mm}
%\end{figure}

\noindent\textbf{Correctness in Java code generation.}
For Java,  we observe that LLMs exhibit different performance patterns than for Python. 
Chinese goes slightly worse for \gp (31.2\% vs. 32.78\%, with a small effect size), and significantly worse (32.30\% vs. 37.13\%, large effect size) for \cl. Hindi goes much worse (17.91\% vs. 37.13\%) with \cl, and the same happens for Spanish, yet with a small effect size (35.00\%  vs 37.13\%). Italian goes better with \ds. In all other combinations, no significant differences were found. This suggests that advantages or performance worsening do not consistently translate across all models and neural architectures, including training-specific methods (\eg Reinforcement Learning with Human Feedback (RLHF) \cite{christiano2017deep}).

\begin{resultbox}
\textbf{RQ$_1$ summary:}  Counterintuitively, prompting in English does not always lead to the best correctness in code generation tasks. Interestingly, we found that prompting in Chinese leads to a consistently better performance \revminor{in Python code generation on CoderEval.} However, none of the languages always produces better results than the others for Java code generation.
\end{resultbox}

\rev{\textbf{Replication on ClassEval.}}
\rev{To assess the generalizability of the CoderEval correctness findings, we replicated the analysis on the ClassEval benchmark~\cite{du2023classeval} (100 classes, 410 methods, Python only), reporting class-level pass rates (a class passes only if all its methods pass in the same iteration).
From our findings, it emerges that GPT achieves near-identical class-level pass rates across all five languages (33\%--38\%) with no significant pairwise differences. DeepSeek follows a similar pattern, with Chinese yielding the highest rate (45\%) but no significant comparisons. Claude 3.5 Sonnet, however, shows English achieving the highest class-level pass rate (12.8\%), with three of the four non-English languages significantly lower: Hindi (10.2\%, $g = 0.25$, large), Spanish (10.3\%, $g = 0.19$, medium), and Italian (8.3\%, $g = 0.46$, large). Chinese (11.3\%) does not reach significance.
At the same time, the Chinese prompt advantage observed in CoderEval does not replicate here, though the broader finding -- that non-English prompts do not systematically hurt correctness -- holds for GPT and DeepSeek. The reversal for Claude suggests that prompt language effects may interact with task granularity.}

\subsection{RQ$_2$ Linguistic Impact on Code Quality}
\label{sub:rq2}

\begin{table}[ht]
\centering
\caption{Python: Comparison with metrics and static analysis tools. Statistically significant $p$-values have $\darkred\Uparrow$ for higher values for the respective language (than for English), and $\darkgreen\Downarrow$ for lower values. Statistically insignificant $p$-values are marked as "-"%\BOWEN{I suggest (1) put higher/lower symbol before p-value, (2) hide all the concrete numbers of p-value, instead, use (*) to indicate if it's significant or not. (3) put all those concrete numbers in the  replication package or supplementary materials. Same comment applies to table 5.}
}
\label{tab:metrics-python}
\begin{tabular}{clcccc}
\toprule
& \textbf{Metric} & \textbf{Chinese} &\textbf{Hindi} & \textbf{Spanish} & \textbf{Italian} \\
\midrule
%GPT
& \nloc & -  & $\darkgreen\Downarrow$ & - & -\\
& \cc & - & $\darkgreen\Downarrow$ & $\darkgreen\Downarrow$ & $\darkgreen\Downarrow$ \\
& \cmt & $\darkred\Uparrow$ & - & $\darkred\Uparrow$ & $\darkgreen\Downarrow$ \\
& \cmd & $\darkred\Uparrow$ & $\darkred\Uparrow$ & - & $\darkgreen\Downarrow$ \\
\textbf{\gp} & \cogn & $\darkgreen\Downarrow$ & $\darkgreen\Downarrow$ & $\darkgreen\Downarrow$ & $\darkgreen\Downarrow$ \\
%& \smells & 0.0558 $\darkred\Uparrow$  & \textbf{0.0282} $\darkred\Uparrow$ & 0.108 $\darkred\Uparrow$  & 0.9139 $\darkred\Uparrow$  \\
& \bugs & - & - & $\darkgreen\Downarrow$ & - \\
%Security.Hotspots & 0.5709 $\darkgreen\Downarrow$  & \textbf{<0.001} $\darkgreen\Downarrow$ & \textbf{<0.001} $\darkgreen\Downarrow$ & 0.0011 $\darkgreen\Downarrow$ * \\
& \pylint & $\darkgreen\Downarrow$ & $\darkred\Uparrow$ & $\darkred\Uparrow$ & $\darkred\Uparrow$ \\
& \flake & $\darkred\Uparrow$ & $\darkred\Uparrow$ & $\darkred\Uparrow$ & $\darkred\Uparrow$ \\
\midrule
%DEEPSEEK
& \nloc & $\darkred\Uparrow$ & $\darkgreen\Downarrow$ & $\darkgreen\Downarrow$ & $\darkgreen\Downarrow$ \\
& \cc & - & $\darkgreen\Downarrow$ & $\darkgreen\Downarrow$ & $\darkgreen\Downarrow$ \\
& \cmt & $\darkred\Uparrow$ & $\darkgreen\Downarrow$ & $\darkgreen\Downarrow$ & $\darkgreen\Downarrow$ \\
& \cmd & $\darkred\Uparrow$ & $\darkgreen\Downarrow$ & $\darkgreen\Downarrow$ & $\darkgreen\Downarrow$ \\
\textbf{\ds} & \cogn & $\darkgreen\Downarrow$ & $\darkgreen\Downarrow$ & $\darkgreen\Downarrow$ & $\darkgreen\Downarrow$ \\
%& \smells & \textbf{0.0091} $\darkred\Uparrow$ & 0.075 $\darkgreen\Downarrow$  & 0.541 $\darkgreen\Downarrow$  & 0.3849 $\darkgreen\Downarrow$  \\
& \bugs & - & - & - & $\darkgreen\Downarrow$ \\
%Security.Hotspots & 0.0144 $\darkgreen\Downarrow$ * & \textbf{<0.001} $\darkgreen\Downarrow$ & 0.0056 $\darkgreen\Downarrow$ * & \textbf{<0.001} $\darkgreen\Downarrow$ \\
& \pylint & $\darkred\Uparrow$ & $\darkred\Uparrow$ & $\darkred\Uparrow$ & $\darkred\Uparrow$ \\
& \flake & $\darkred\Uparrow$ & $\darkred\Uparrow$ & $\darkred\Uparrow$ & $\darkred\Uparrow$ \\
\midrule
%CLAUDE
& \nloc & $\darkred\Uparrow$ & $\darkgreen\Downarrow$ & $\darkgreen\Downarrow$ & - \\
& \cc & $\darkred\Uparrow$ & - & $\darkgreen\Downarrow$ & $\darkgreen\Downarrow$ \\
& \cmt & $\darkred\Uparrow$ & $\darkgreen\Downarrow$ & $\darkgreen\Downarrow$ & $\darkgreen\Downarrow$ \\
& \cmd & $\darkred\Uparrow$ & $\darkgreen\Downarrow$ & $\darkgreen\Downarrow$ & $\darkgreen\Downarrow$ \\
\textbf{\cl} & \cogn & $\darkred\Uparrow$ & $\darkred\Uparrow$ & - & $\darkgreen\Downarrow$ \\
%& \smells & \textbf{0.0044} $\darkgreen\Downarrow$ & 0.7351 $\darkgreen\Downarrow$  & \textbf{0.009} $\darkgreen\Downarrow$ & \textbf{<0.001} $\darkgreen\Downarrow$ \\
& \bugs & $\darkred\Uparrow$ & - & - & $\darkgreen\Downarrow$ \\
%Security.Hotspots & 0.5082 $\darkred\Uparrow$  & \textbf{<0.001} $\darkgreen\Downarrow$ & 0.1317 $\darkgreen\Downarrow$  & 0.2813 $\darkgreen\Downarrow$  \\
& \pylint & $\darkred\Uparrow$ & $\darkred\Uparrow$ & $\darkred\Uparrow$ & $\darkred\Uparrow$ \\
& \flake & $\darkred\Uparrow$ & $\darkred\Uparrow$ & $\darkred\Uparrow$ & $\darkred\Uparrow$ \\
\bottomrule
\end{tabular}
\end{table}

\begin{table}[ht]
\centering
\caption{Java: Comparison with metrics and static analysis tools. Stat. signif. $p$-values have $\darkred\Uparrow$ for higher values for the respective language (than for English), and $\darkgreen\Downarrow$ for lower values. Not Stat. signif. values are marked as "-"}
\label{tab:metrics-java}
\begin{tabular}{clcccc}
\toprule
& \textbf{Metric} & \textbf{Chinese} &\textbf{Hindi} & \textbf{Spanish} & \textbf{Italian} \\
\midrule
%GPT
& \nloc & $\darkgreen\Downarrow$ & $\darkgreen\Downarrow$ & - & - \\
& \cc & - & $\darkgreen\Downarrow$ & - & - \\
& \cmt & - & - & $\darkred\Uparrow$ & - \\
\textbf{\gp} & \cmd & - & $\darkred\Uparrow$ & - & - \\
& \cogn & - & $\darkgreen\Downarrow$ & - & - \\
& \smells & - & $\darkgreen\Downarrow$ & $\darkgreen\Downarrow$ & - \\
& \bugs & - & $\darkgreen\Downarrow$ & - & - \\
%Security.Hotspots & 0.0545 $\darkred\Uparrow$  & \textbf{<0.001} $\darkred\Uparrow$ & \textbf{<0.001} $\darkred\Uparrow$ & 0.1502 $\darkred\Uparrow$  \\
& \pmd & - & - & - & - \\
\midrule
%DEEPSEEK
& \nloc & $\darkred\Uparrow$ & $\darkred\Uparrow$ & $\darkred\Uparrow$ & $\darkred\Uparrow$ \\
& \cc & $\darkred\Uparrow$ & - & - & $\darkred\Uparrow$ \\
& \cmt & $\darkred\Uparrow$ & - & - & - \\
\textbf{\ds} & \cmd & $\darkgreen\Downarrow$ & $\darkgreen\Downarrow$ & - & $\darkgreen\Downarrow$ \\
& \cogn & $\darkred\Uparrow$ & - & $\darkred\Uparrow$ & $\darkred\Uparrow$ \\
& \smells & $\darkred\Uparrow$ & - & $\darkred\Uparrow$ & $\darkred\Uparrow$ \\
& \bugs & - & - & $\darkgreen\Downarrow$ & - \\
%Security.Hotspots & 0.0247 $\darkred\Uparrow$ * & 0.9182 $\darkgreen\Downarrow$  & \textbf{<0.001} $\darkred\Uparrow$ & 0.0153 $\darkred\Uparrow$ * \\
& \pmd & - & - & - & - \\
\midrule
%CLAUDE
& \nloc & $\darkred\Uparrow$ & - & - & - \\
& \cc & $\darkred\Uparrow$ & $\darkred\Uparrow$ & - & - \\
& \cmt & $\darkred\Uparrow$ & $\darkgreen\Downarrow$ & $\darkgreen\Downarrow$ & $\darkgreen\Downarrow$ \\
\textbf{\cl} & \cmd & - & $\darkgreen\Downarrow$ & - & - \\
& \cogn & $\darkred\Uparrow$ & $\darkred\Uparrow$ & - & - \\
& \smells & $\darkgreen\Downarrow$ & $\darkgreen\Downarrow$ & $\darkgreen\Downarrow$ & $\darkgreen\Downarrow$ \\
& \bugs & - & - & $\darkgreen\Downarrow$ & $\darkgreen\Downarrow$ \\
%Security.Hotspots & \textbf{<0.001} $\darkred\Uparrow$ & 0.0313 $\darkgreen\Downarrow$ * & 0.0042 $\darkgreen\Downarrow$ * & nan $\darkgreen\Downarrow$ * \\
& \pmd & $\darkgreen\Downarrow$ & - & $\darkgreen\Downarrow$ & $\darkgreen\Downarrow$ \\
\bottomrule
\end{tabular}
\end{table}

\tabref{tab:metrics-python} and \tabref{tab:metrics-java} report the results of the comparison between English and other languages in terms of metrics and number of static analysis tools' warnings (described in Section~\ref{sec:codequality}) in Python and Java datasets, respectively. \rev{All metrics and static analysis indicators were computed independently on each of the ten generated samples per task. Statistical comparisons (Wilcoxon signed-rank tests) were performed on the resulting paired distributions across all iterations, ensuring that our findings are robust to cross-iteration variability.} In particular, \tabref{tab:warnings-python} and \tabref{tab:warnings-java} detail the results of the comparison between English and other languages in terms of the category of warnings and errors reported by the static tools (these categories are described in \tabref{tab:errordescription}). We use Wilcoxon signed-rank $p$-values (statistically significant differences between English and the given language are shown in boldface) because our data is paired in different languages and minimal assumptions about data distribution. We further indicate the impact of these differences using ($\darkred\Uparrow$) when the value is greater for the non-English language, while reverting the arrow ($\darkgreen\Downarrow$) in the other cases. Effect sizes, less relevant in this context, are available in our replication package.
\rev{To support a more nuanced interpretation, the severity levels introduced in \tabref{tab:errordescription} should be considered when assessing the practical significance of the observed differences. An increase in minor-severity warnings (\eg formatting or style issues) has different maintenance implications than an increase in critical-severity issues (\eg runtime faults or security risks). We organize the following discussion accordingly, highlighting when severity levels differ across the observed patterns.}

\begin{figure*}[htbp]
\centering
\includegraphics[width=0.9\columnwidth]{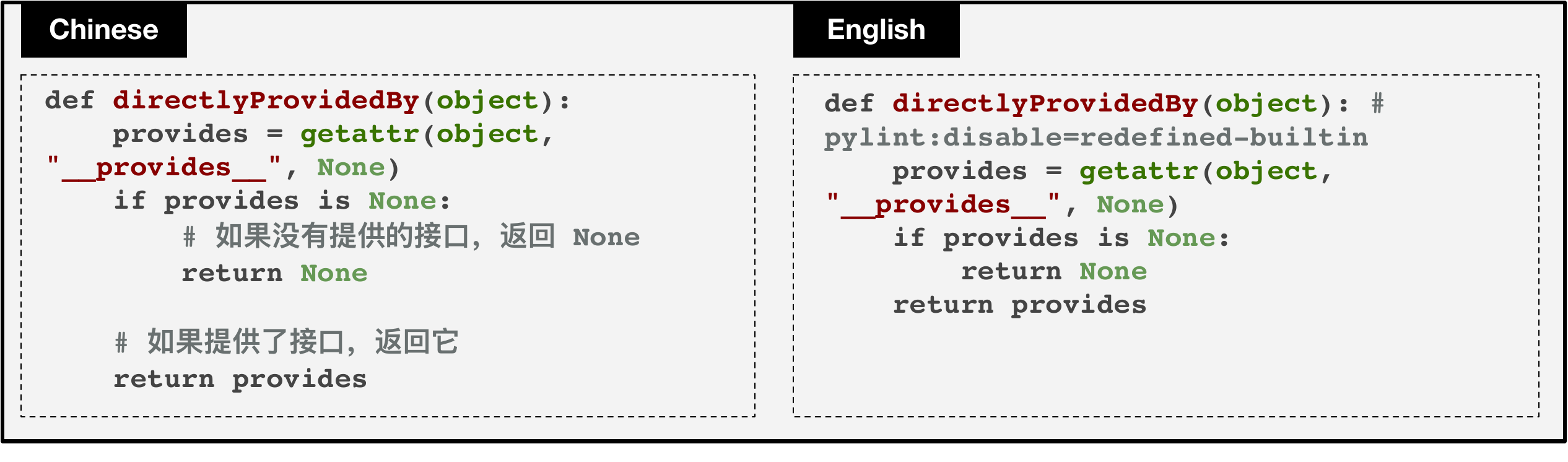}
\caption{Generated code from different language prompts: Chinese (left), and English (right). The Chinese code includes more comments.}
\label{fig:listing2}
\end{figure*}

\noindent\textbf{Analysis in Python code generation.}
For Python (\tabref{tab:metrics-python}), the most striking observation is the presence of significant differences across multiple metrics and tool-generated warnings.
%even if, in most cases, the effect size of such differences is negligible. However, as explained in \secref{}, even a negligible (yet significant) difference means more lines of code, comments, and more decision points, \ie a different code, which is what we wanted to observe.
 \gp recommended \textbf{code has lower cyclomatic complexity (except for Chinese), and is also shorter for Hindi}.
%\BOWEN{pls mention the concrete numbers here so that reviewers can easily connect to the table} 
\rev{\ds shows the same cyclomatic-complexity pattern, but its code is also shorter than English for Spanish and Italian, and notably \emph{longer} for Chinese.}
\rev{For \ds and \cl, \textbf{Chinese prompts yield more comments than English} (10.4\% vs.\ 5.3\% for \ds; 17.6\% vs.\ 11.9\% for \cl), while for \gp the two are nearly equivalent (8.7\% vs.\ 9.1\%).} 
%While we cannot establish a definitive cause, this pattern may reflect stylistic conventions in the Chinese-language training corpora these models have been exposed to,} confirming the analysis provided above detailing the prevailing growth of China in data-intensive domains (\secref{sub:rq1}).
While we cannot establish a definitive cause, this pattern may reflect stylistic conventions in the Chinese-language training corpora these models have been exposed to, \revminor{though, as noted in \secref{sub:rq1}, such interpretations remain observational rather than confirmed causes.}
\figref{fig:listing2} depicts an example illustrating the prevalence of comments in Python code generated from a Chinese prompt. On the left is the code generated with a Chinese prompt; on the right is the code generated using an English prompt. Although the code looks identical, the Chinese version contains more comments describing the two conditional branches of the \texttt{if} statement.
Hindi only increases comments' density for \gp, while it decreases for the other models. Finally, for \ds and \cl, Spanish and Italian prompts exhibit the same behavior in recommending code that, on average, includes fewer comments.
The cognitive complexity always decreases for \gp, \ds, and \cl-Italian, while it increases for \cl-Chinese and \cl-Spanish.  %Smells, depending on the model and language, either increase or decrease and, except for \cl, rarely present significant differences. 
\sonar smells increase for \gp-Hindi and \ds-Chinese, while they always decrease for \cl except for Hindi.  Bugs decrease for \gp-Spanish, \ds-Italian, and \cl-Italian, while they increase for \cl-Chinese. 

\pylint warnings consistently show a significant increase across all languages, except for code generated by \gp with Chinese prompts. Similarly, \flake exhibits a rise in warnings across all languages and LLMs. In particular, for \gp-Hindi/Spanish/Italian, \ds-Hindi, and \cl-Hindi, the effect size remains within the small to medium range (see replication package). \rev{However, a closer inspection of the warning categories (\tabref{tab:warnings-python}) reveals that these increases are predominantly driven by minor-severity issues, particularly PycodestyleWarnings and convention violations related to formatting. Critical-severity warnings (\eg \pylint fatal/error and \flake PyFlakesErrors) show no consistent increase for non-English prompts, suggesting that while non-English prompts may trigger more stylistic deviations, they do not systematically introduce more severe defects.} For the warnings, we noticed this is mainly due to strange formatting when generating code in these languages.

% \noindent
% \begin{minipage}[t]{0.48\textwidth}
% \begin{lstlisting}[style=mypython, caption={}, label={}]
% def directlyProvidedBy(object):
%     provides = getattr(object, "__provides__", None)
%     if provides is None:
%         # 如果没有提供的接口，返回 None
%         return None
    
%     # 如果提供了接口，返回它
%     return provides
% \end{lstlisting}
% \end{minipage}%
% \hfill
% \begin{minipage}[t]{0.48\textwidth}
% \begin{lstlisting}[style=mypython, caption={}, label={}]
% def directlyProvidedBy(object): # pylint:disable=redefined-builtin
%     provides = getattr(object, "__provides__", None)
%     if provides is None:
%         return None
%     return provides
% \end{lstlisting}
% \end{minipage}

% \captionof{figure}{Generated code from different language prompts: Chinese (left), and English (right). The Chinese code includes more comments.}
% \label{fig:listing2}

%\vspace{-10pt}
\setlength{\textfloatsep}{10pt}

\begin{table}[ht]
\centering
\caption{Python: Comparison with static analysis tools errors and warnings. Statistically significant $p$-values have $\darkred\Uparrow$ for higher values for the respective language (than for English), and $\darkgreen\Downarrow$ for lower values. Statistically  insignificant $p$-values are marked as "-"}
\label{tab:warnings-python}
\vspace{-2mm}
\scriptsize
\resizebox{0.7\textwidth}{!}{%
\begin{tabular}{cclcccc}
\toprule
\textbf{Model} & \textbf{Tool} & \textbf{Metric} & \textbf{Chinese} &\textbf{Hindi} & \textbf{Spanish} & \textbf{Italian} \\
\midrule
%GPT
  & \flake & BugRisks.BestPractices & $\darkgreen\Downarrow$  & -  & $\darkgreen\Downarrow$  & $\darkgreen\Downarrow$  \\
 & \flake & DocstringIssues & $\darkred\Uparrow$  & $\darkred\Uparrow$  & $\darkgreen\Downarrow$  & $\darkgreen\Downarrow$  \\
 & \flake & McCabeComplexity & $\darkgreen\Downarrow$  & -  & -  & -  \\
 & \flake & PyFlakesErrors & -  & -  & -  & $\darkgreen\Downarrow$  \\
 & \flake & PycodestyleErrors & $\darkgreen\Downarrow$  & $\darkred\Uparrow$  & $\darkred\Uparrow$  & $\darkred\Uparrow$  \\
 & \flake & PycodestyleWarnings & $\darkred\Uparrow$  & $\darkred\Uparrow$  & $\darkred\Uparrow$  & $\darkred\Uparrow$  \\
 & \pylint & fatal & -   & -   & -   & -  \\
 & \pylint & error & -  & $\darkgreen\Downarrow$  & -  & $\darkgreen\Downarrow$  \\
 & \pylint & warning & $\darkred\Uparrow$  & -  & $\darkred\Uparrow$  & -  \\
 \textbf{\gp}& \pylint & convention & -  & $\darkred\Uparrow$  & $\darkred\Uparrow$  & $\darkred\Uparrow$  \\
 & \pylint & refactor & $\darkred\Uparrow$  & -  & -  & -  \\
 & \sonar & BUG.BLOCKER & $\darkgreen\Downarrow$  & -   & $\darkgreen\Downarrow$  & $\darkgreen\Downarrow$  \\
 & \sonar & BUG.MAJOR & -  & $\darkred\Uparrow$  & $\darkgreen\Downarrow$  & $\darkgreen\Downarrow$  \\
 & \sonar & BUG.MINOR & -   & $\darkgreen\Downarrow$  & -   & -   \\
 & \sonar & CODE.SMELL.BLOCKER & -  & -   & $\darkgreen\Downarrow$  & $\darkgreen\Downarrow$  \\
 & \sonar & CODE.SMELL.CRITICAL & $\darkgreen\Downarrow$  & $\darkred\Uparrow$  & $\darkgreen\Downarrow$  & $\darkgreen\Downarrow$  \\
 & \sonar & CODE.SMELL.MAJOR & $\darkgreen\Downarrow$  & $\darkred\Uparrow$  & $\darkgreen\Downarrow$  & $\darkgreen\Downarrow$  \\
 & \sonar & CODE.SMELL.MINOR & $\darkgreen\Downarrow$  & $\darkred\Uparrow$  & $\darkgreen\Downarrow$  & $\darkgreen\Downarrow$  \\
 & \sonar & SecurityHotspot & -  & $\darkgreen\Downarrow$  & $\darkgreen\Downarrow$  & $\darkgreen\Downarrow$  \\
\midrule
%DEEPSEEK
 & \flake & BugRisks.BestPractices & $\darkred\Uparrow$  & $\darkred\Uparrow$  & $\darkred\Uparrow$  & -  \\
 & \flake & DocstringIssues & $\darkred\Uparrow$  & $\darkred\Uparrow$  & $\darkgreen\Downarrow$  & $\darkgreen\Downarrow$  \\
 & \flake & McCabeComplexity & -  & $\darkgreen\Downarrow$  & $\darkgreen\Downarrow$  & $\darkgreen\Downarrow$  \\
 & \flake & PyFlakesErrors & -  & $\darkgreen\Downarrow$  & $\darkgreen\Downarrow$  & -  \\
 & \flake & PycodestyleErrors & $\darkgreen\Downarrow$  & $\darkred\Uparrow$  & $\darkred\Uparrow$  & $\darkred\Uparrow$  \\
 & \flake & PycodestyleWarnings & $\darkred\Uparrow$  & $\darkred\Uparrow$  & $\darkred\Uparrow$  & $\darkred\Uparrow$  \\
 & \pylint & fatal & -   & -   & -   & -   \\
 & \pylint & error & $\darkred\Uparrow$  & -  & -  & -  \\
 & \pylint & warning & $\darkred\Uparrow$  & -  & $\darkred\Uparrow$  & $\darkred\Uparrow$  \\
 \textbf{\ds}& \pylint & convention & $\darkred\Uparrow$  & $\darkred\Uparrow$  & $\darkred\Uparrow$  & $\darkred\Uparrow$  \\
 & \pylint & refactor & $\darkred\Uparrow$  & -  & -  & -  \\
 & \sonar & BUG.BLOCKER & $\darkgreen\Downarrow$  & $\darkgreen\Downarrow$  & $\darkgreen\Downarrow$  & $\darkgreen\Downarrow$  \\
 & \sonar & BUG.MAJOR & $\darkgreen\Downarrow$  & $\darkred\Uparrow$  & $\darkgreen\Downarrow$  & $\darkgreen\Downarrow$  \\
 & \sonar & BUG.MINOR & -   & -   & -   & -   \\
 & \sonar & CODE.SMELL.BLOCKER & -   & -   & -   & -   \\
 & \sonar & CODE.SMELL.CRITICAL & $\darkgreen\Downarrow$  & $\darkred\Uparrow$  & $\darkgreen\Downarrow$  & $\darkgreen\Downarrow$  \\
 & \sonar & CODE.SMELL.MAJOR & $\darkgreen\Downarrow$  & $\darkred\Uparrow$  & $\darkgreen\Downarrow$  & $\darkgreen\Downarrow$  \\
 & \sonar & CODE.SMELL.MINOR & $\darkgreen\Downarrow$  & $\darkred\Uparrow$  & $\darkgreen\Downarrow$  & $\darkgreen\Downarrow$  \\
 & \sonar & SecurityHotspot & $\darkgreen\Downarrow$  & $\darkgreen\Downarrow$  & $\darkgreen\Downarrow$  & $\darkgreen\Downarrow$  \\
\midrule
% CLAUDE
 & \flake & BugRisks.BestPractices & $\darkgreen\Downarrow$  & -  & $\darkgreen\Downarrow$  & -  \\
 & \flake & DocstringIssues & $\darkred\Uparrow$  & $\darkred\Uparrow$  & $\darkgreen\Downarrow$  & $\darkgreen\Downarrow$  \\
 & \flake & McCabeComplexity & -  & -  & -  & -  \\
 & \flake & PyFlakesErrors & -  & -  & $\darkgreen\Downarrow$  & -  \\
 & \flake & PycodestyleErrors & $\darkgreen\Downarrow$  & $\darkred\Uparrow$  & $\darkred\Uparrow$  & $\darkred\Uparrow$  \\
 & \flake & PycodestyleWarnings & $\darkred\Uparrow$  & $\darkred\Uparrow$  & $\darkred\Uparrow$  & $\darkred\Uparrow$  \\
 & \pylint & fatal & -   & -   & -   & -   \\
 & \pylint & error & -  & $\darkred\Uparrow$  & $\darkgreen\Downarrow$  & $\darkred\Uparrow$  \\
 & \pylint & warning & $\darkgreen\Downarrow$  & -  & -  & -  \\
 \textbf{\cl}& \pylint & convention & $\darkred\Uparrow$  & $\darkred\Uparrow$  & $\darkred\Uparrow$  & $\darkred\Uparrow$  \\
 & \pylint & refactor & $\darkgreen\Downarrow$  & -  & $\darkgreen\Downarrow$  & $\darkgreen\Downarrow$  \\
 & \sonar & BUG.BLOCKER & -  & $\darkgreen\Downarrow$  & $\darkgreen\Downarrow$  & $\darkgreen\Downarrow$  \\
 & \sonar & BUG.MAJOR & $\darkgreen\Downarrow$  & $\darkgreen\Downarrow$  & $\darkgreen\Downarrow$  & $\darkgreen\Downarrow$  \\
 & \sonar & BUG.MINOR & $\darkgreen\Downarrow$  & -  & $\darkgreen\Downarrow$  & $\darkgreen\Downarrow$  \\
 & \sonar & CODE.SMELL.BLOCKER & $\darkgreen\Downarrow$  & -   & $\darkgreen\Downarrow$  & $\darkgreen\Downarrow$  \\
 & \sonar & CODE.SMELL.CRITICAL & $\darkgreen\Downarrow$  & $\darkred\Uparrow$  & $\darkgreen\Downarrow$  & $\darkgreen\Downarrow$  \\
 & \sonar & CODE.SMELL.MAJOR & $\darkgreen\Downarrow$  & $\darkred\Uparrow$  & $\darkgreen\Downarrow$  & $\darkgreen\Downarrow$  \\
 & \sonar & CODE.SMELL.MINOR & $\darkgreen\Downarrow$  & $\darkred\Uparrow$  & $\darkgreen\Downarrow$  & $\darkgreen\Downarrow$  \\
 & \sonar & SecurityHotspot & -  & $\darkgreen\Downarrow$  & -  & -  \\
\bottomrule
\end{tabular}
}
\end{table}

Regarding the warning categories (\tabref{tab:warnings-python}), several interesting patterns emerge. Starting from \flake metrics, for \gp, \textbf{significant decreases occur in all non-English languages except Hindi}. \ds shows increases for Chinese, Hindi, and Spanish, suggesting the generated code violates best practices more in these languages. \cl mostly shows decreases, especially for Chinese and Spanish, indicating that code best practices are inconsistently affected by language and model, with \ds producing more risky code in non-English prompts. The \texttt{McCabeComplexity} mostly decreases or is non-significant, except for \ds in Hindi/Spanish; indeed, \textbf{non-English prompts tend to produce slightly simpler functions}, though differences are generally small. Moving to \pylint metrics, it occurs a uniform trend among convention, warning and refactor metrics: \gp shows mixed effects: convention issues increase in Hindi, Spanish, Italian, but decrease in Chinese, while \ds and \cl convention violations increase in almost all non-English languages, highlighting \textbf{LLMs may produce code that deviates more from PEP8 conventions when prompted in non-English languages}, except in some isolated cases (Chinese for \gp). For Error and Fatal metrics, some decreases in Spanish and Italian, but generally non-English prompts do not strongly increase runtime-critical errors, suggesting syntax and execution correctness are largely preserved. Finally, for \sonar metrics, \textbf{bugs generally decrease for Chinese, Spanish, Italian} (\gp and \ds), \textbf{but increase for Hindi} in some cases. Hindi prompts may yield slightly riskier code, while other languages are generally safer in terms of critical bugs. For code smells, across all models, code smells are mostly decreased for Chinese, Spanish, and Italian but increased for Hindi, indicating structural or stylistic issues may worsen specifically for Hindi prompts, but improve or remain neutral for others.

\begin{figure*}[htbp]
\centering
\includegraphics[width=0.9\columnwidth]{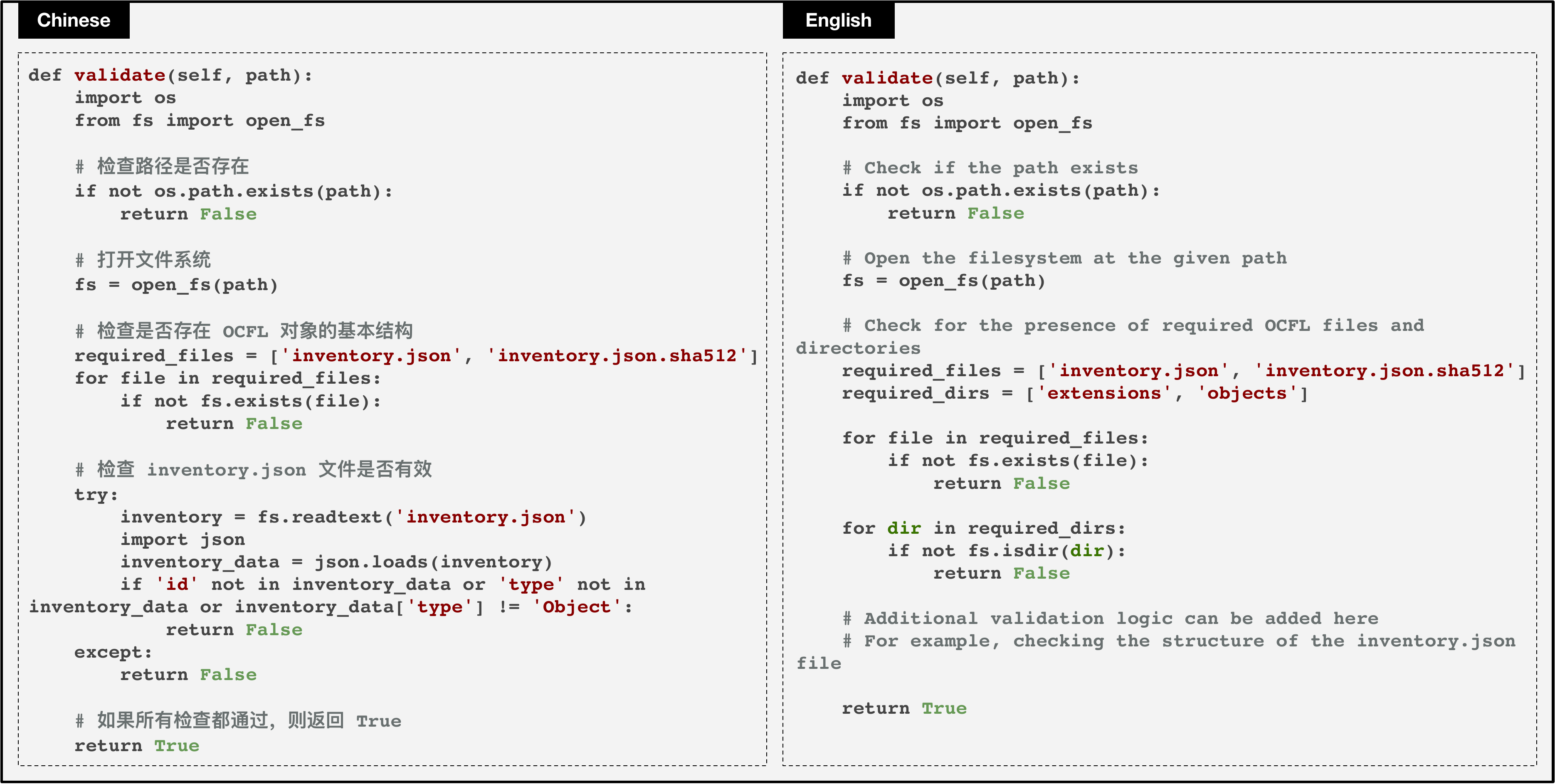}
\caption{Generated code from Chinese (left), and English (right). The two codes look completely different despite the task being the same, in particular, the Chinese version presents a \flake best practices warning related to the except statement.}
\label{fig:bestpractices}
\end{figure*}

%\BOWEN{it's a long paragraph, pls bold the key findings so that readers can quickly capture them.}
\figref{fig:bestpractices} shows a comparison between a code generated with a Chinese prompt (left) and English prompt (right). The two codes look different, despite the task to implement being the same. The English code checks only the structural presence of required files and directories. It does not care whether \texttt{inventory.json} is valid JSON or has the right fields, whereas the Chinese code checks only that required files exist and that \texttt{inventory.json} is valid JSON with specific fields. It does not check for the required directories. \rev{This example demonstrates how prompt language can influence the model's interpretation of task requirements: the Chinese-generated code prioritizes data validation (checking JSON structure and required fields), while the English-generated code prioritizes file system completeness (checking for required directories). Both represent valid but distinct implementation strategies, each with its own trade-offs in terms of robustness and coverage.} Moreover, the Chinese version has been reported by \flake as violating the best practice B0001, \ie ``Do not use bare `except:`, it also catches unexpected events like memory errors, interrupts, system exit, and so on.  Prefer excepting specific exceptions  If you're sure what you're doing, be explicit and write `except BaseException''.

\begin{figure*}[htbp]
\centering
\includegraphics[width=0.9\columnwidth]{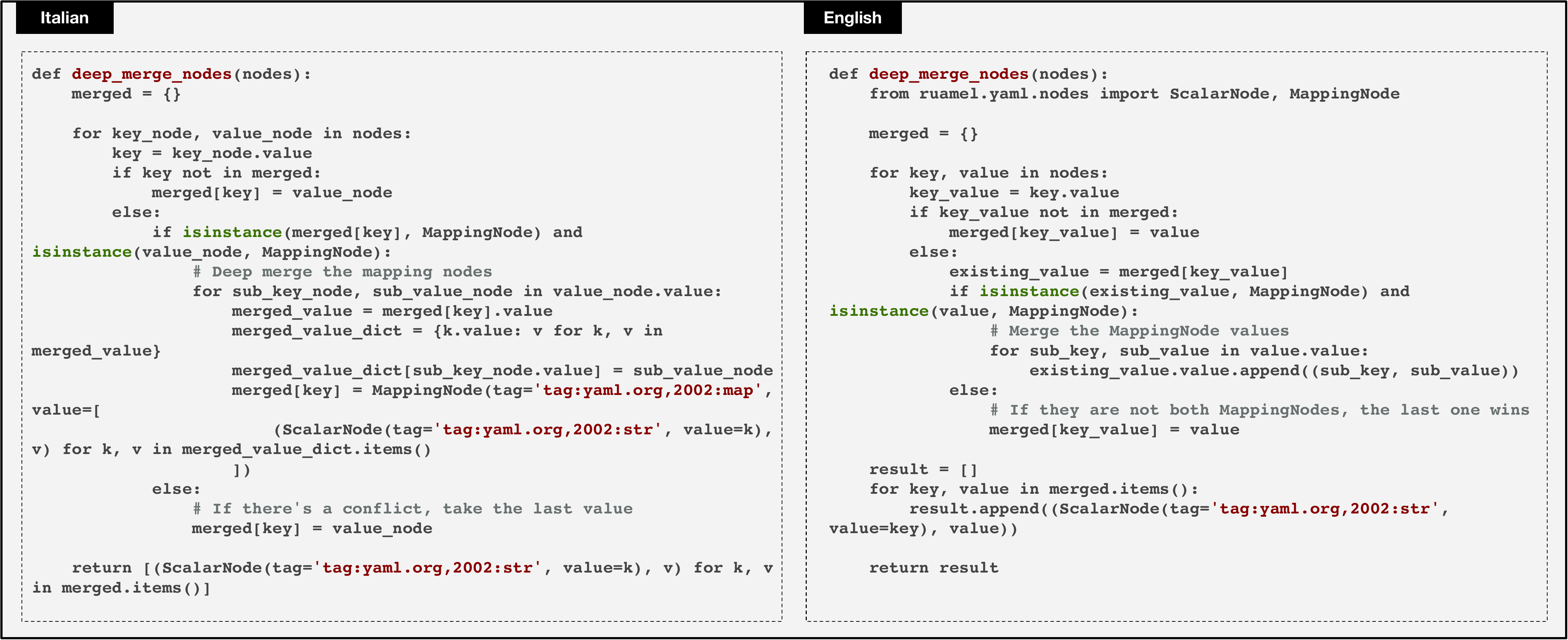}
\caption{Generated code from Italian (left), and English (right). The two codes exhibit differences in terms of \pylint convention warning reported (eight warnings for the Italian code against five for the English code)}
\label{fig:convention}
\end{figure*}
%\vspace{-10pt}
\setlength{\textfloatsep}{10pt}

%\BOWEN{replace all the double quotation marks "" with ``''} \ALESSANDRO{FIXED}
\figref{fig:convention} displays a comparison between a code generated with an Italian prompt (left) and an English prompt (right). In this case, the Italian code exhibits eight convention warnings reported by \pylint compared to the five reported for the English code. In particular, four ``line-too-long'' warnings, ``trailing-whitespace'', ``missing-final-newline'', `` missing-module-docstring'' and ``invalid-name'' warnings were reported for the Italian, while ``missing-final-newline'', ``missing-module-docstring'', ``invalid-name'', ``missing-function-docstring'' and ``import-outside-toplevel'' warnings for the english code. \rev{Notably, the majority of the Italian warnings (six out of eight) are minor-severity formatting issues (``line-too-long,'' ``trailing-whitespace,'' ``missing-final-newline''), whereas the English warnings include moderate-severity organizational concerns such as ``import-outside-toplevel.'' While the Italian code triggers more warnings in absolute count, the English code exhibits issues with greater potential impact on maintainability.} The Italian generation style was denser (long lines, inline constructs, quick variable names), leading to more formatting-related warnings; conversely, the English generation style was cleaner in formatting, but lacked documentation and had non-standard import placement, leading to more ``organization'' warnings.

\begin{figure*}[h!]
\centering
\includegraphics[width=0.9\columnwidth]{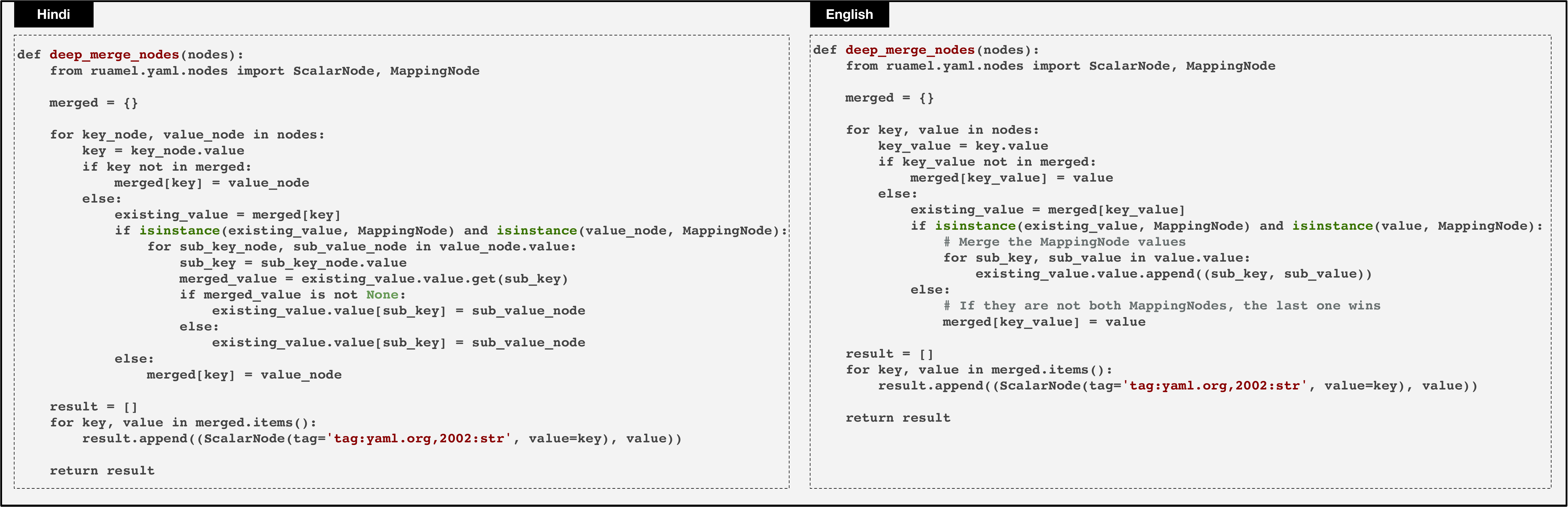}
\caption{Generated code from Hindi (left), and English (right). The two codes exhibit differences in terms of \sonar major bug reported (one bug for the Hindi code against zero for the English code)}
\label{fig:bug}
\end{figure*}
%\vspace{-10pt}
\setlength{\textfloatsep}{10pt}

Lastly, \figref{fig:bug} shows the comparison between a code generated with a Hindi prompt and an English prompt. In this scenario, the difference is related to the number of major bugs reported by \sonar. \textbf{While the English code is free from major bugs, the Hindi code exhibits a major bug} on lines 16-18: the if and else statement executes the exact same assignment, hence adding complexity without changing the outcome.
Sonar flags it as a ``bug'' because it looks like you intended different behavior depending on the condition, but forgot to implement it.
\rev{This pattern, where Hindi-generated code introduces redundant control flow without functional impact, was observed in multiple instances and contributes to the higher bug counts reported for Hindi across our statistical analysis. It suggests that prompts in Hindi may lead the model toward more verbose but logically incomplete implementations, potentially reflecting differences in how the model processes Hindi-language specifications.}

\noindent\textbf{Metrics in Java code generation.}
For Java, what we observe is slightly different than for Python. 
\rev{Before discussing the metrics, it is worth noting that the static analysis results reported in this section are computed only on successfully compiled Java code. As shown in \tabref{tab:tests}, compilation rates for GPT and DeepSeek are generally stable across prompt languages, indicating that the exclusion of non-compiling samples does not introduce a language-dependent bias. Claude-Hindi, however, exhibits a notably lower compilation rate (33.91\% vs. 46.61\%--50.87\% for other languages), and the findings for this combination should therefore be interpreted in light of this reduced sample.}
%More specifically, in Java code generation, we never observed non-negligible effect sizes, not even for static analysis tools for which we report aggregated results (\pmd).
If we look at \gp, Chinese and Italian, in general, do not exhibit significant differences from English (except for longer code for Chinese), while  Hindi leads to less code and, consequently, a decrease in most of the other indicators (see the $\darkgreen\Downarrow$ near the \nloc metric, as well as near the other indicators). Spanish exhibits more comments and fewer \textsc{SonarCloud} smells.
For \ds, we observe longer code (NLOC) in all cases, and also more complex (CC) for Chinese and Italian. We also see an increase in cognitive complexity and \textsc{SonarCloud} Smells (except for Hindi). We do not observe any significant difference in terms of \textsc{SonarCloud} bugs (except for a reduction in the case of Spanish) and for \pmd warnings.
Regarding \cl, code length increases only for Chinese, while complexity rises for both Chinese and Hindi. Although Chinese prompts lead to code with a higher number of comments, the opposite trend is observed for the other languages. Cognitive Complexity shows a significant increase for Chinese and Hindi, whereas the overall count of code smells decreases.  The number of detected bugs declines when using Spanish and Italian prompts. Except for Hindi, \pmd warnings exhibit a significant reduction.

\begin{figure*}[htbp]
\centering
\includegraphics[width=0.9\columnwidth]{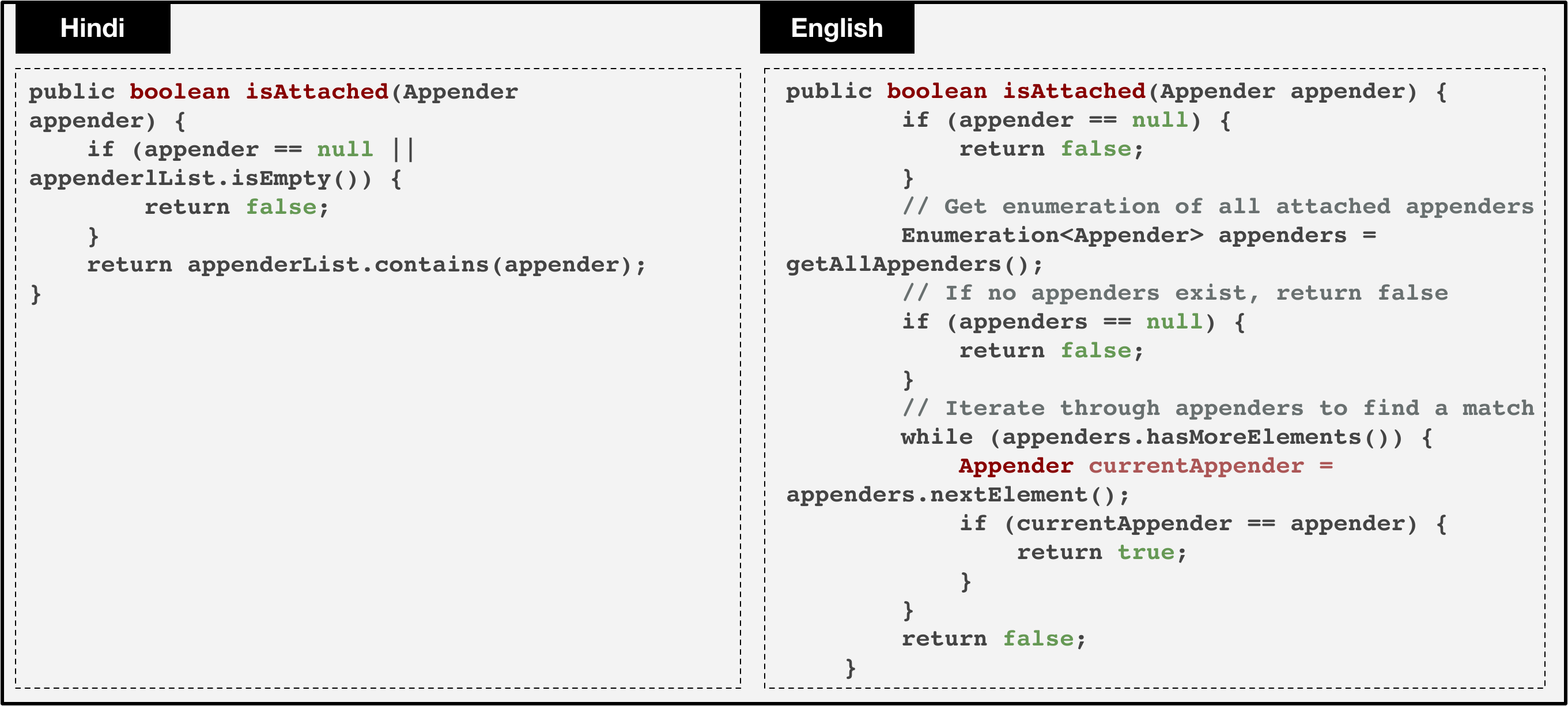}
\caption{Generated code from Hindi (left), and English (right). The English code has a higher cyclomatic complexity than Hindi.}
\label{fig:listing3}
\end{figure*}
%\vspace{-10pt}
\setlength{\textfloatsep}{10pt}

Fig.~\ref{fig:listing3} illustrates an example of differences in cyclomatic complexity between two Java code snippets generated using prompts in different languages. On the left side, the code is generated using a Hindi prompt, while on the right side, the code is generated using an English prompt. %This example highlights a notable distinction between the two versions: 
While the English-generated code contains more \texttt{if} and \texttt{while} statements, thereby increasing its cyclomatic complexity, the Hindi-generated code uses just a single \texttt{if} statement to implement the functionality. Despite these structural differences, both code versions successfully pass all tests in the benchmark. However, the English-generated code performs additional checks that, while likely not covered by the test suite, may enhance the code's robustness.
%\BOWEN{highlight/bold the key message at the end we want to send based on this example}\ALESSANDRO{FIXED}

\begin{table}[ht]
\centering
\caption{Java: Comparison with static analysis tools errors and warnings. Statistically significant $p$-values have $\darkred\Uparrow$ for higher values for the respective language (than for English), and $\darkgreen\Downarrow$ for lower values. Statistically insignificant $p$.values are marked as "-"}
\label{tab:warnings-java}
\scriptsize
\resizebox{0.65\textwidth}{!}{%
\begin{tabular}{cclcccc}
\toprule
\textbf{Model} & \textbf{Tool} & \textbf{Metric} & \textbf{Chinese} &\textbf{Hindi} & \textbf{Spanish} & \textbf{Italian} \\
\midrule
%GPT
 & \pmd & best.practices & -  & -  & $\darkgreen\Downarrow$  & -  \\
 & \pmd & codestyle & -  & $\darkred\Uparrow$  & $\darkred\Uparrow$  & $\darkred\Uparrow$  \\
 & \pmd & design & -  & $\darkgreen\Downarrow$  & -  & $\darkgreen\Downarrow$  \\
 & \pmd & documentation & $\darkgreen\Downarrow$  & $\darkgreen\Downarrow$  & $\darkred\Uparrow$  & $\darkred\Uparrow$  \\
 & \pmd & error.prone & -  & $\darkgreen\Downarrow$  & $\darkgreen\Downarrow$  & -  \\
 & \pmd & multithreading & $\darkgreen\Downarrow$  & $\darkgreen\Downarrow$  & $\darkgreen\Downarrow$  & $\darkgreen\Downarrow$  \\
 & \pmd & performance & -  & $\darkgreen\Downarrow$  & -  & -  \\
 & \sonar & BUG.BLOCKER & -   & -   & -   & -   \\
 \textbf{\gp} & \sonar & BUG.CRITICAL & -   & -   & -   & -   \\
 & \sonar & BUG.MAJOR & -   & -   & -   & $\darkgreen\Downarrow$  \\
 & \sonar & BUG.MINOR & -  & $\darkgreen\Downarrow$  & $\darkgreen\Downarrow$  & $\darkgreen\Downarrow$  \\
 & \sonar & CODE.SMELL.BLOCKER & -   & -   & -   & -   \\
 & \sonar & CODE.SMELL.CRITICAL & $\darkgreen\Downarrow$  & $\darkgreen\Downarrow$  & $\darkgreen\Downarrow$  & $\darkgreen\Downarrow$  \\
 & \sonar & CODE.SMELL.MAJOR & $\darkred\Uparrow$  & $\darkgreen\Downarrow$  & $\darkgreen\Downarrow$  & $\darkgreen\Downarrow$  \\
 & \sonar & CODE.SMELL.MINOR & $\darkgreen\Downarrow$  & $\darkgreen\Downarrow$  & $\darkgreen\Downarrow$  & $\darkgreen\Downarrow$  \\
 & \sonar & CODE.SMELL.INFO & $\darkgreen\Downarrow$  & $\darkgreen\Downarrow$  & $\darkgreen\Downarrow$  & $\darkgreen\Downarrow$  \\
 & \sonar & SecurityHotspot & $\darkred\Uparrow$  & $\darkred\Uparrow$  & $\darkred\Uparrow$  & -  \\
\midrule
%DEEPSEEK
  & \pmd & best.practices & $\darkred\Uparrow$  & -  & $\darkred\Uparrow$  & $\darkred\Uparrow$  \\
 & \pmd & codestyle & $\darkred\Uparrow$  & $\darkgreen\Downarrow$  & $\darkred\Uparrow$  & $\darkred\Uparrow$  \\
 & \pmd & design & $\darkred\Uparrow$  & $\darkgreen\Downarrow$  & $\darkred\Uparrow$  & $\darkred\Uparrow$  \\
 & \pmd & documentation & $\darkgreen\Downarrow$  & $\darkgreen\Downarrow$  & $\darkred\Uparrow$  & $\darkred\Uparrow$  \\
 & \pmd & error.prone & $\darkred\Uparrow$  & -  & -  & $\darkred\Uparrow$  \\
 & \pmd & multithreading & -  & -  & $\darkgreen\Downarrow$  & -  \\
 & \pmd & performance & -  & -  & -  & $\darkred\Uparrow$  \\
 & \sonar & BUG.BLOCKER & $\darkgreen\Downarrow$  & -  & -   & -  \\
 \textbf{\ds} & \sonar & BUG.CRITICAL & -   & -  & -   & -   \\
 & \sonar & BUG.MAJOR & $\darkgreen\Downarrow$  & $\darkgreen\Downarrow$  & $\darkgreen\Downarrow$  & $\darkgreen\Downarrow$  \\
 & \sonar & BUG.MINOR & $\darkgreen\Downarrow$  & $\darkgreen\Downarrow$  & $\darkgreen\Downarrow$  & $\darkgreen\Downarrow$  \\
 & \sonar & CODE.SMELL.BLOCKER & -   & -  & -   & -   \\
 & \sonar & CODE.SMELL.CRITICAL & $\darkgreen\Downarrow$  & $\darkgreen\Downarrow$  & $\darkgreen\Downarrow$  & $\darkgreen\Downarrow$  \\
 & \sonar & CODE.SMELL.MAJOR & $\darkred\Uparrow$  & $\darkgreen\Downarrow$  & $\darkgreen\Downarrow$  & $\darkgreen\Downarrow$  \\
 & \sonar & CODE.SMELL.MINOR & $\darkgreen\Downarrow$  & $\darkgreen\Downarrow$  & $\darkgreen\Downarrow$  & $\darkgreen\Downarrow$  \\
 & \sonar & CODE.SMELL.INFO & -  & $\darkgreen\Downarrow$  & -   & -   \\
 & \sonar & SecurityHotspot & $\darkred\Uparrow$  & -  & $\darkred\Uparrow$  & $\darkred\Uparrow$  \\
\midrule
% CLAUDE
  & \pmd & best.practices & -  & $\darkgreen\Downarrow$  & -  & -  \\
 & \pmd & codestyle & -  & $\darkred\Uparrow$  & $\darkgreen\Downarrow$  & -  \\
 & \pmd & design & -  & $\darkred\Uparrow$  & $\darkred\Uparrow$  & -  \\
 & \pmd & documentation & $\darkgreen\Downarrow$  & $\darkgreen\Downarrow$  & $\darkred\Uparrow$  & $\darkred\Uparrow$  \\
 & \pmd & error.prone & $\darkgreen\Downarrow$  & $\darkgreen\Downarrow$  & $\darkgreen\Downarrow$  & $\darkgreen\Downarrow$  \\
 & \pmd & multithreading & $\darkred\Uparrow$  & $\darkred\Uparrow$  & $\darkgreen\Downarrow$  & -  \\
 & \pmd & performance & -  & -  & -  & -  \\
 & \sonar & BUG.BLOCKER & -   & -   & -   & -   \\
 \textbf{\cl} & \sonar & BUG.CRITICAL & -   & -  & $\darkgreen\Downarrow$  & $\darkgreen\Downarrow$  \\
 & \sonar & BUG.MAJOR & -   & $\darkgreen\Downarrow$  & $\darkgreen\Downarrow$  & $\darkgreen\Downarrow$  \\
 & \sonar & BUG.MINOR & -   & -   & -   & -   \\
 & \sonar & CODE.SMELL.BLOCKER & -   & -   & -   & -   \\
 & \sonar & CODE.SMELL.CRITICAL & $\darkgreen\Downarrow$  & $\darkgreen\Downarrow$  & $\darkgreen\Downarrow$  & $\darkgreen\Downarrow$  \\
 & \sonar & CODE.SMELL.MAJOR & $\darkgreen\Downarrow$  & $\darkgreen\Downarrow$  & $\darkgreen\Downarrow$  & $\darkgreen\Downarrow$  \\
 & \sonar & CODE.SMELL.MINOR & $\darkgreen\Downarrow$  & $\darkgreen\Downarrow$  & $\darkgreen\Downarrow$  & $\darkgreen\Downarrow$  \\
 & \sonar & CODE.SMELL.INFO & -   & -   & -  & $\darkgreen\Downarrow$  \\
 & \sonar & SecurityHotspot & $\darkred\Uparrow$  & $\darkgreen\Downarrow$  & $\darkgreen\Downarrow$  & -  \\
\bottomrule
\end{tabular}
}
\end{table}

Discussing the warning categories (\tabref{tab:warnings-java}), interesting patterns appear from the analysis. Starting from PMD
%\BOWEN{Instead of combine everything in one long paragraph, I suggest to use itemize format to elaborate each patten. And use one sentence at the beginning to summarize the pattern. For example, \newline ``$\bullet$ \textbf{[PATTERN SUMMARY].} [DETAILED DESCRIPTION AND SUPPORT].''\newline}
\textbf{best practices}, \gp shows decreases in Spanish and Chinese, but Hindi is mostly unaffected, \ds often shows increases for Chinese, Spanish, and Italian, indicating more violations of best practices, while \cl mostly shows decreases, particularly for Chinese and Spanish, concluding that best-practice compliance is inconsistently affected by prompt language and model, with \ds producing slightly more risky code in non-English prompts. \textbf{code style and design issues}: \gp increases for Hindi, Spanish, and Italian for code style, while design issues decrease for Hindi and Italian. \ds Mostly increases for CodeStyle and Design in Chinese, Spanish, and Italian, while \cl shows mixed metrics: some decreases for Chinese and Spanish, while Hindi shows slight increases. Style and design adherence is sensitive to both model and prompt language, with non-English prompts often producing more style warnings. \textbf{multithreading}:  consistently lower for \gp and \ds, while mixed for \cl. Error-prone issues decreased in Hindi and Spanish for \gp and \cl, and mixed in \ds. \textbf{Functional correctness and performance-related}: less affected by prompt language, though multithreading concerns are consistently mitigated by \gp and \ds.

\begin{figure*}[htbp]
\centering
\includegraphics[width=0.9\columnwidth]{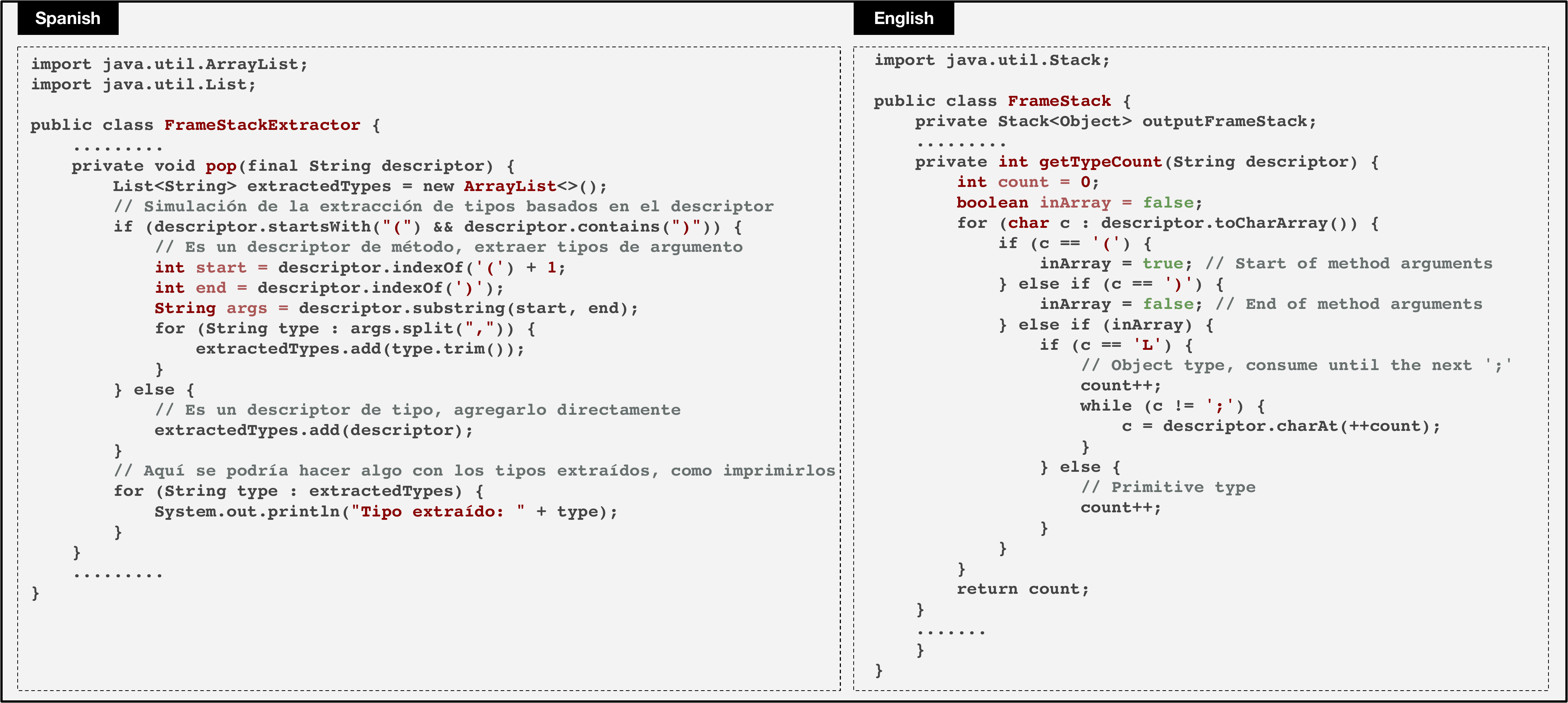}
\caption{Generated code from Spanish (left), and English (right). The two codes exhibits differences in terms of \pmd error prone warnings reported (zero for the Spanish code against three for the english code).}
\label{fig:pmderror}
\end{figure*}
%\vspace{-10pt}
\setlength{\textfloatsep}{10pt}

%\BOWEN{I'd suggest moving each example around the corresponding context in the analysis paragraph. Instead of describing them separately.}\ALESSANDRO{FIXED}
\figref{fig:pmderror} depicts a comparison between a code generated with a Spanish prompt (left) and an English prompt (right). \textbf{The English code exhibits three 'error prone' warnings detected by PMD, while for the Spanish version none was detected}. In particular, the English code presents three ``AvoidLiteralsInIfCondition'', \ie avoid using hard-coded literals in conditional statements. By declaring them as static variables or private members with descriptive names, maintainability is enhanced. 

Moving to \sonar, minor and major bugs generally decrease for Chinese, Spanish, and Italian (\gp and \ds), but Hindi sometimes increases, showing that Hindi prompts may yield slightly riskier code, while other languages are safer in terms of critical bugs. Across all models, code smells are mostly decreased for Chinese, Spanish, and Italian, but increase for Hindi. Structural or stylistic issues worsen specifically for Hindi prompts, but improve or remain neutral for others.

\begin{resultbox}
\textbf{RQ$_2$ summary:}  
\revminor{Across the size, complexity, and comment metrics and the static-analysis warnings from \flake, \pylint, \pmd, and \sonar examined in this study---and, for Java, on the subset of outputs that successfully compile---prompts in languages other than English do not necessarily lead to lower code quality.} However, the observed differences, especially for Python, suggest that the generated code varies in structure or implementation details. In particular, non-English prompts tend to produce simpler code but more style/convention violations, with Hindi often yielding riskier outputs (more bugs and smells), while Chinese, Spanish, and Italian generally show improvements or remain stable.

%Static analysis warnings and metrics indicate that prompts in languages other than English do not necessarily lead to lower code quality. However, the observed differences, especially for Python, suggest that the generated code varies in structure or implementation details. In particular, non-English prompts tend to produce simpler code but more style/convention violations, with Hindi often yielding riskier outputs (more bugs and smells), while Chinese, Spanish, and Italian generally show improvements or remain stable.

%different code is being generated.\BOWEN{The latter sentence is unclear, what does ``different code'' mean?\ALESSANDRO{FIXED}}
\end{resultbox}

\rev{\noindent\textbf{Replication on ClassEval.}
For ClassEval, all Wilcoxon signed-rank tests for Claude yield negligible Cliff's $\delta$ values (ranging from $-0.11$ to $0.07$), confirming that prompt language does not produce practically meaningful differences in code structure or static analysis outcomes. Chinese prompts yield slightly fewer \flake warnings ($\delta = -0.11$) and \pylint warnings ($\delta = -0.01$), while Hindi, Spanish, and Italian yield slightly more -- a pattern consistent with CoderEval. The descriptive statistics for GPT and DeepSeek (from the single iteration described in Section~3.1) show the same trend: mean NLOC, CCN, \flake, and \pylint values are comparable across all prompt languages, with no systematic differences attributable to prompt language. Because all Cliff's $\delta$ values are negligible and the GPT and DeepSeek per-language means differ by less than one unit on every indicator, we do not extend Tables~5--8 with a ClassEval block; the full per-language means and statistical test results are available in the replication package. These results confirm that the structural similarity of generated code across prompt languages, observed in CoderEval, generalizes to class-level tasks.}

\begin{table}[ht]
\caption{\rev{Analysis of source-code comments and string literals across benchmarks.} Values are percentages (\%) of all comments or string literals that contain text in the indicated language(s).
E = English, C = Chinese, H = Hindi, S = Spanish, I = Italian, B = both the target language and English.}
%\BOWEN{What does the number represent? I'd suggest briefly explaining in the caption to make the table self-contained.\ALESSANDRO{FIXED}}
\label{tab:lexicon}
    \centering
    \resizebox{\linewidth}{!}{
    \begin{tabular}{cl|rrrrrr|rrrrrr|rrrrrr|rrrrrr}
    \toprule
    & & \multicolumn{6}{c|}{\textbf{Chinese}} &\multicolumn{6}{c|}{\textbf{Hindi}} & \multicolumn{6}{c|}{\textbf{Spanish}} & \multicolumn{6}{c}{\textbf{Italian}} \\ 
    & & \multicolumn{3}{c}{\bf Comments (\%)} & \multicolumn{3}{c|}{\bf Literals (\%)} & \multicolumn{3}{c}{\bf Comments (\%)} & \multicolumn{3}{c|}{\bf Literals (\%)} & \multicolumn{3}{c}{\bf Comments (\%)} & \multicolumn{3}{c|}{\bf Literals (\%)} & \multicolumn{3}{c}{\bf Comments (\%)} & \multicolumn{3}{c}{\bf Literals (\%)} \\
    & \textbf{Model} & \textbf{E} & \textbf{C} & \textbf{B} & \textbf{E} & \textbf{C} & \textbf{B} & \textbf{E} & \textbf{H} & \textbf{B} & \textbf{E} & \textbf{H} & \textbf{B} & \textbf{E} & \textbf{S} & \textbf{B} & \textbf{E} & \textbf{S} & \textbf{B} & \textbf{E} & \textbf{I} & \textbf{B} & \textbf{E} & \textbf{I} & \textbf{B}\\
    \midrule
    %Python
    & \gp & \rev{9.1} & 8.7 & 0.6 & 11.0 & 3.0 & 0.0 & 15.3 & 2.8 & 0.0 & 6.2 & 1.1 & 0.8 & 8.3 & 10.6 & 0.2 & 15.9 & 6.4 & 0.9 & 11.7 & 7.0 & 0.2 & 9.3 & 5.1 & 0.0 \\
    \textbf{Python} & \ds & 5.3 & 10.4 & 2.5 & 12.3 & 0.9 & 0.8 & 14.9 & 2.3 & 0.2 & 4.7 & 1.7 & 0.4 & 6.0 & 11.7 & 0.0 & 11.1 & 7.4 & 0.9 & 11.5 & 7.6 & 0.0 & 7.9 & 5.1 & 0.0 \\
& \cl & 11.9 & 17.6 & 5.3 & 16.1 & 0.8 & 0.6 & 30.6 & 0.9 & 0.0 & 7.0 & 0.4 & 0.4 & 13.8 & 15.1 & 0.0 & 16.1 & 2.8 & 2.1 & 21.5 & 9.26 & 0.0 & 8.9 & 4.0 & 0.0 \\
\midrule
%Java
& \gp & 27.6 & 3.6 & 1.9 & 21.4 & 1.7 & 0.6 & 31.4 & 1.5 & 0.0 & 14.4 & 0.6 & 0.0 & 19.5 & 10.2 & 0.4 & 19.7 & 7.2 & 0.2 & 19.8 & 3.0 & 0.2 & 14.9 & 2.3 & 0.0 \\
\textbf{Java} & \ds & 11.7 & 10.2 & 3.97 & 18.1 & 0.9 & 0.6 & 26.3 & 1.9 & 0.0 & 13.8 & 0.76 & 0.2 & 12.3 & 15.1 & 1.1 & 12.1 & 6.8 & 0.6 & 10.4 & 11.0 & 0.0 & 8.5 & 4.0 & 0.4 \\
& \cl & 17.8 & 7.0 & 0.6 & 6.8 & 0.2 & 0.0 & 20.6 & 0.0 & 0.0 & 7.0 & 0.0 & 0.0 & 13.8 & 4.2 & 0.2 & 5.5 & 1.1 & 0.0 & 18.1 & 3.0 & 0.0 & 5.29 & 0.4 & 0.0 \\
\midrule
%ClassEval (NEW)
& \rev{\gp} & \rev{0.2} & \rev{0.0} & \rev{0.0} & \rev{2.2} & \rev{0.2} & \rev{0.0} & \rev{0.0} & \rev{0.0} & \rev{0.0} & \rev{5.6} & \rev{1.0} & \rev{0.0} & \rev{0.0} & \rev{0.0} & \rev{0.0} & \rev{30.2} & \rev{2.7} & \rev{0.2} & \rev{0.0} & \rev{0.0} & \rev{0.0} & \rev{37.1} & \rev{2.2} & \rev{0.0} \\
\rev{\textbf{ClassEval}} & \rev{\ds} & \rev{0.2} & \rev{0.2} & \rev{0.0} & \rev{5.4} & \rev{0.5} & \rev{0.0} & \rev{0.0} & \rev{0.0} & \rev{0.0} & \rev{8.5} & \rev{1.2} & \rev{0.0} & \rev{0.2} & \rev{0.0} & \rev{0.0} & \rev{36.1} & \rev{2.9} & \rev{0.2} & \rev{0.0} & \rev{0.0} & \rev{0.0} & \rev{32.7} & \rev{2.9} & \rev{0.0} \\
& \rev{\cl} & \rev{19.0} & \rev{25.6} & \rev{0.7} & \rev{4.2} & \rev{1.5} & \rev{0.0} & \rev{40.5} & \rev{3.4} & \rev{0.0} & \rev{6.1} & \rev{1.9} & \rev{0.0} & \rev{23.7} & \rev{16.8} & \rev{0.2} & \rev{35.4} & \rev{4.2} & \rev{0.0} & \rev{0.0} & \rev{0.0} & \rev{0.0} & \rev{34.9} & \rev{5.6} & \rev{0.0} \\
    \bottomrule
    \end{tabular}
    }
\end{table}

\subsection{RQ$_3$ Linguistic Impact on Code Lexicon} \label{sub:rq3}

\tabref{tab:lexicon} reports the results of the manual analysis for comments and string literals. We do not report results for identifiers because, except for extremely rare exceptions (detailed in the replication package), as expected, they were only generated in English.
%\MARIO{Guys, I remember that in Spanish, we found few cases of identifiers in both languages and only in Spanish (not in the manual validation sample, but in the 460 instances sample) }. 
For comments and string literals, the table shows the percentage of cases in which they were generated in English (E), specific language (C/H/S/I), or both (B).

\noindent\textbf{Code lexicon in Python code generation.}
For Python, comments exhibit two contrasting patterns. Chinese and Spanish have a larger proportion of comments in their respective languages than English comments across all models except for \gp in Chinese, where the percentages are similar. Instead, Hindi and Italian predominantly feature English comments across all the models.
For literals, a consistent behavior appears among languages and models, with a preference for English literals over those in other languages. \cl shows a higher consistency in using the same language, always showing the lowest values of `both' occurrences. \gp and \ds tend to blend languages when writing comments or literals, showing inconsistency even if in a small portion of cases. A notable difference arises in Italian, where most occurrences do not mix languages, except for \gp with comments.

\noindent\textbf{Code lexicon in Java code generation.}
For Java, comments reveal a stable trend among languages and models, with a preference for English. \rev{Notable exceptions exist in Spanish and Italian with the \ds model, where the proportion of comments in the target language exceeds that of English (Spanish: 15.1\% vs. 12.3\%; Italian: 11.0\% vs. 10.4\%), indicating a shift away from the general English-dominant trend.} For literals, a notable prevalence of English literals spans across languages and models. 
%Finally, 'both' occurrences reflect the same trend as in Python, where Chinese and Spanish typically show more mixed languages in comments and literals compared to Hindi and Italian. 
As far as cases where comments or literals appear in both (`b') languages, \cl shows a better consistency in using the same language than the other LLMs.

\begin{table}[ht]
\centering
\caption{Percentage of cases in which comments and string literals are in inconsistent languages (C: Chinese, H: Hindi, S: Spanish, I: Italian). }
\label{tab:inconsistency}

\resizebox{0.4\textwidth}{!}{%
\begin{tabular}{clrrrr}
\toprule
& \textbf{Model} & \textbf{C (\%)} & \textbf{H (\%)} & \textbf{S (\%)} & \textbf{I (\%)} \\ 
\midrule
% Python
& \gp & 8 & 1 & 8 & 3 \\
\textbf{Python} & \ds & 10 & 1 & 8 & 3 \\
& \cl & 19 & 1 & 16 & 3 \\
\midrule
% Java
& \gp & 7 & 2 & 11 & 3 \\
\textbf{Java} & \ds & 15 & 1 & 9 & 3 \\
& \cl & 1 & 0 & 1 & 0 \\
\bottomrule
\end{tabular}%
}

%\vspace{-3mm}
\end{table}

\noindent\textbf{Inconsistencies between comments and string literals.}
\tabref{tab:inconsistency} reports the percentage of cases where comments and literals are written in two different languages, \eg comments are in English and literals in Chinese.

For Python, inconsistencies are more prevalent in Chinese and Spanish than in Italian and Hindi, where their occurrence is notably lower.
%This pattern reinforces the tendency of these models to favor English and, in this case, to mix it up frequently with their output in other languages. 
%Considerably, Italian and Hindi present fewer cases where languages are mixed, confirming the previous analysis where the English language is often preferred for both comments and literals. 
On the LLMs' side--\cl appears to be the worst of the three models, with the highest percentage of inconsistencies for Chinese and Spanish.

Java, Chinese, and Spanish were confirmed to exhibit more mixups than other languages, with the highest percentage for both \gp and \ds. However, \cl shows an opposite trend to that of Python, exhibiting an almost null percentage of mixed cases.

\rev{Comparing the inconsistency patterns across Python and Java (\tabref{tab:inconsistency}) reveals notable differences. In Python, \cl exhibits the highest inconsistency rates (19\% for Chinese, 16\% for Spanish), whereas in Java, \cl achieves near-zero inconsistency (1\% for Chinese, 0\% for Hindi, 1\% for Spanish, 0\% for Italian). This suggests that \cl handles lexicon consistency differently depending on the target programming language, potentially reflecting differences in the model's training data distribution across Python and Java codebases. In contrast, \gp and \ds show more consistent inconsistency patterns across both programming languages, with Chinese and Spanish exhibiting higher mixup rates than Hindi and Italian in both cases.}

We found that \textbf{regardless of the prompt language, LLMs generate identifiers in English}, perhaps because of the provided signature, and in particular for Chinese and Hindi, because programming languages do not support such characters to name the identifiers.
%\BOWEN{For chinese prompt, I think it's because the PLs do not support use chinese characters to name the identifiers. So english is the only option regardless the signatures.}
This is in line with previous work showing how function signatures drive LLMs~\cite{ding2024code} for generating code summaries. Furthermore, developers tend to write identifiers in English rather than their native language~\cite{pawelkacode2015}.

To further investigate the effect of signatures contained in the prompt, as explained in \secref{sec:rq3method} we conducted a controlled experiment on 100 randomly selected Python examples for which signatures have been translated into Italian.

\begin{table}[ht]
\centering
\caption{Given the language requested on the prompt (\ie Italian or English), the values represent the percentage of cases in which comments, identifiers and string literals are in Italian (I), English (E), Both (B) or None(N).}
\label{tab:ablation}
\resizebox{\linewidth}{!}{
\begin{tabular}{l|rrrrrrrrrrrr|rrrrrrrrrrrr}
\toprule
&  \multicolumn{12}{c|}{\textbf{Italian}} & \multicolumn{12}{c}{\textbf{English}} \\
& \multicolumn{4}{c}{\textbf{Comments(\%)}} & \multicolumn{4}{c}{\textbf{Identifiers(\%)}} & \multicolumn{4}{c|}{\textbf{Literals(\%)}} & \multicolumn{4}{c}{\textbf{Comments(\%)}} & \multicolumn{4}{c}{\textbf{Identifiers(\%)}} & \multicolumn{4}{c}{\textbf{Literals(\%)}} \\
\multicolumn{1}{c|}{\textbf{Model}} & \multicolumn{1}{c}{\textbf{I}} & \multicolumn{1}{c}{\textbf{E}} & \multicolumn{1}{c}{\textbf{B}} & \multicolumn{1}{c}{\textbf{N}} & \multicolumn{1}{c}{\textbf{I}} & \multicolumn{1}{c}{\textbf{E}} & \multicolumn{1}{c}{\textbf{B}} & \multicolumn{1}{c}{\textbf{N}} & \multicolumn{1}{c}{\textbf{I}} & \multicolumn{1}{c}{\textbf{E}} & \multicolumn{1}{c}{\textbf{B}} & \multicolumn{1}{c|}{\textbf{N}} & \multicolumn{1}{c}{\textbf{I}} & \multicolumn{1}{c}{\textbf{E}} & \multicolumn{1}{c}{\textbf{B}} & \multicolumn{1}{c}{\textbf{N}} & \multicolumn{1}{c}{\textbf{I}} & \multicolumn{1}{c}{\textbf{E}} & \multicolumn{1}{c}{\textbf{B}} & \multicolumn{1}{c}{\textbf{N}} & \multicolumn{1}{c}{\textbf{I}} & \multicolumn{1}{c}{\textbf{E}} & \multicolumn{1}{c}{\textbf{B}} & \multicolumn{1}{c}{\textbf{N}} \\
\midrule
\gp & 24.00 & 0.00 & 0.00 & 76.00 & 88.00 & 0.00 & 10.00 & 2.00 & 28.00 & 2.00 & 2.00 & 68.00 & 35.29 & 0.00 & 0.00 & 64.71 & 29.41 & 23.53 & 41.18 & 5.88 & 23.53 & 0.00 & 0.00 & 76.47 \\
\ds & 26.00 & 0.00 & 0.00 & 74.00 & 92.00 & 0.00 & 8.00 & 0.00 & 46.00 & 0.00 & 0.00 & 54.00 & 90.20 & 0.00 & 0.00 & 9.80 & 17.65 & 27.45 & 52.94 & 1.96 & 25.49 & 0.00 & 0.00 & 74.51 \\
\cl & 72.00 & 14.00 & 0.00 & 14.00 & 36.00 & 6.00 & 58.00 & 0.00 & 36.00 & 14.00 & 2.00 & 48.00 & 100.00 & 0.00 & 0.00 & 0.00 & 13.73 & 9.80 & 72.55 & 3.92 & 21.57 & 1.96 & 1.96 & 74.51 \\
\bottomrule
\end{tabular}
}
%\vspace{-3mm}
\end{table}

\tabref{tab:ablation} depicts the results of this controlled experiment study. The column \texttt{Italian} represent the cases where the signature is translated in Italian, while the \texttt{English} column where the signature is kept in english. The single columns shows the cases when Comments, Identifiers or Literals are respectively in Italian (I), English (E), both languages (B) or no elements found for that entry (N), \ie there are codes without comments or literals or even identifier (\eg the method contains just method calls). 

We found that--\textbf{when provided with an Italian method signature, \gp and \ds predominantly produced comments, identifiers, and literals in Italian} (with just 2\% of literals in English for \gp, and 0\% for \ds), with only rare cases of mixed-language identifiers (10\% and 8\% for \gp and \ds). These exceptions typically involved technical terms that are challenging to translate (\eg ``wrapper,''  ``directory,'' or ``output''). Both \gp and \ds consistently generated comments in Italian, with occasional English literals limited to default messages. In contrast, \cl exhibited less consistency, frequently mixing English and Italian for identifiers (58\%), while less for literals (2\%).

With English signatures and explicit instructions for Italian output, all three models produced identifiers with greater variability, even though comments and literals consistently appeared in Italian. \gp (41.18\% of the case) and \ds (52.94\%) sometimes translated the signature despite the prompt's instruction to retain it, leading to the use of Italian. When the signature remained in English, these LLMs produced a mix of Italian and English identifiers—sometimes preserving English attributes from the method signature, other times translating them inconsistently. In this scenario, \cl showed consistency by never translating the signature, though it still generated mixed-language identifiers (72.55\%).

While the language used for a function signature in the prompts may partially drive identifiers' language, LLMs do not guarantee a consistent code lexicon when asked to provide specifications in a given natural language. 

\begin{resultbox}
%\textbf{RQ$_3$ summary:} Our manual analysis of comments and literals highlights a notable inconsistency when generating literals and comments. Models tend to indiscriminately use English regardless of the language used in the prompt. Yet, the presence of extreme cases is not to be exluded--for which the recommendation features  a of mix multiple languages within the same code.
%sometimes

\textbf{RQ$_3$ summary:} 
 We found a notable inconsistency in LLMs when generating literals
and comments. LLMs tend to indiscriminately use English
regardless of the language used in the prompt and sometimes
blend multiple languages within the same code.
Overall, function signature language partially drives output, but LLMs do not guarantee a consistent code lexicon in the target language.
\end{resultbox}

\rev{\noindent\textbf{Replication on ClassEval.}
The ClassEval rows in \tabref{tab:lexicon} report the lexicon analysis. GPT and DeepSeek generate almost no comments on ClassEval ($>$99\% of methods contain no comments), limiting the lexicon analysis for these models to literals only. Claude, however, shows patterns consistent with CoderEval: Chinese (25.6\%) and Spanish (16.8\%) prompts produce comments in the target language, while Hindi predominantly yields English comments (40.5\%) and Italian generates no comments. For literals, all three models exhibit a similar pattern: Spanish and Italian prompts produce the highest proportion of English-language literals (30--37\%), while Chinese and Hindi produce fewer ($<$9\%). Identifiers are generated in English in 97--100\% of cases across all model--language combinations, consistent with CoderEval. These findings confirm that LLMs exhibit inconsistent lexicon behavior across prompt languages, though the degree of inconsistency varies with the benchmark and model.}

\section{Implications of our Findings}
\label{sec:implications}

In the following, we summarize the implications of our findings, targeting practitioners and educators.

% mainly targeting practitioners and, secondarily, creators of LLMs for code. 
%\MAX{This is just an attempt. I really need help with further ideas.. feel free to reshape it completely!!}

\subsection{Implications for Practitioners}

\textbf{Specifications can be left in their own native language, but$\dots$} As stated in the introduction, there are several scenarios in which developers start from specifications written in their language and maybe want/have to keep them as part of the code documentation (\ie docstrings/Javadoc). 
Our results indicate that when specifications are written in a language other than English, code correctness does not necessarily decrease. 
%However, this may happen for  certain combinations of LLMs and programming languages (\eg \cl and Java). This means that, 
In general, the choice of avoiding translating specifications is viable, yet before making such a choice, developers should try to understand whether their specific programming language, domain, and the chosen LLM allow doing so.

\rev{Our findings suggest specific guidelines: for Python development, Chinese prompts may actually yield higher correctness (RQ$_1$), while for Java, the choice of prompt language should be made with greater caution, as some combinations---particularly Claude with Hindi---exhibit substantially lower compilation rates and functional correctness. Developers should therefore conduct pilot evaluations on a representative sample of their tasks before committing to non-English prompts in production workflows.}

\textbf{The source code quality is not worse. It is just a different code.}
We found that source code quality indicators do not indicate that code quality, \eg in terms of maintainability or readability indicators, is necessarily worse. Sometimes, there are precautions to be adopted, \eg prompts in given languages may lead to strange formatting that requires an (easy) fix. At the same time, the fact that such indicators are significantly different means that code can be implemented by making different choices or by following different strategies. Also, we noticed that the level of documentation (comments) may often differ.
We have no means to understand so, but this could be based on what the LLM has observed, \eg associated with text in given languages. Future work should investigate the reasons for that better.

\rev{Our severity-aware analysis (RQ$_2$) reveals an important nuance: while non-English prompts tend to increase the total count of static analysis warnings, these increases are predominantly driven by minor-severity style and formatting issues rather than critical defects. In practice, this means that the additional warnings triggered by non-English prompts are largely addressable through automated formatters (\eg \texttt{black} for Python, \texttt{google-java-format} for Java) and do not indicate deeper quality degradation. Developers using non-English prompts should therefore incorporate automated formatting as a standard post-generation step.}

\textbf{One cannot assume that the code will be documented in the same language used for the prompt.} While, on the one hand, as found in our ablation study, the language used for the signature may drive the identifiers' language, on the other hand, comments and literals tend to be very inconsistent in general. Therefore, additional work from the developers' side is needed to ensure their consistency.

\rev{This has practical consequences for teams operating in non-English-speaking environments. If consistent documentation in the team's native language is required, developers should not rely on the LLM's default behavior. Instead, they should either (i) translate the function signature into the target language, which our ablation study shows significantly increases lexicon consistency for GPT and DeepSeek, or (ii) explicitly instruct the model to use the desired language for comments and identifiers in the system prompt. Even with these measures, post-generation review of comments and literals remains advisable, as no configuration fully guarantees lexicon consistency.}

\subsection{Implications for Programming Educators}
Our experimental results demonstrate that the language barrier for learning programming has been broken by LLM. On the one hand, it is certainly a technical leap that enables educators to easily teach how to program. For example, our RQ$_1$ results indicate that whenever educators think of a specification in the class, they can easily come up with a prompt in their own native language for LLM to generate the code without worrying about the performance drop. They can explain to the students the connections between the prompt and the code. This is not the case in the old days. For example, Xu \etal developed a cross-linguistic tool to specifically help Chinese developers translate their technical queries into English~\cite{xu2016domain}. Then, they use the translation to retrieve the relevant knowledge from the online developer forums, such as Stack Overflow. Similar efforts to break the language barriers in the software engineering community can also be found~\cite{chen2016learning,wang2019domain}.
Moreover, the superior performance of LLMs also brings challenges to programming educators. For example, ensuring that students truly understand the knowledge rather than simply using LLM-generated code has become a critical challenge. Our results (especially RQ$_1$ and RQ$_2$) further indicated that even if the students can only speak their native language, they will be able to generate code with decent quality. In other words, the difficulties of generating code are becoming even lower. In summary, \textbf{our overall results show that prompts in different languages lead to insignificant code quality differences that could carry both positive and negative impacts on programming education in this specific code generation context.}

\rev{For educators specifically, our findings carry three actionable implications. First, instructors can confidently use non-English specifications in classroom settings without fearing a significant drop in the quality of LLM-generated code, broadening accessibility for students in non-English-speaking environments. Second, the observed differences in implementation strategies across prompt languages (as illustrated in Figs.~1 and~3) provide natural teaching opportunities: instructors can use these contrasting solutions to discuss algorithmic trade-offs, defensive programming, and the importance of code review. Third, the lexicon inconsistencies documented in RQ$_3$ serve as a concrete example of LLM limitations that students should be aware of, reinforcing the message that LLM-generated code requires human review and should not be blindly trusted.}

\begin{comment}
\subsection{Implications for LLM creators}
Undoubtedly, LLMs are trained in the information available on the Internet. As far as source code is concerned, their internationalization ability largely depends on the amount of source code and development discussions (\eg on SO and other forums) in given languages. The results we obtained for Python and Chinese---largely due to the increasing availability of Chinese code and discussions for such a language---constitute a clear example.
Anybody who wants to train (or fine-tune) a model able to work in a given nationalization environment should carefully target proper training resources. \MAX{is it too trivial/obvious? feel free to refine}
\end{comment}

\section{Threats to Validity} \label{sec:threats}

Threats to \textbf{\emph{construct validity}} concern the relationship between theory and observation. This concerns the extent to which the test cases used could be a valid indicator to address RQ$_1$. While we leveraged test cases from a widely used benchmark, such test cases may not be adequate. This threat is mitigated because, at least, we perform a \emph{relative} assessment, using the same test cases for different languages.
Concerning RQ$_2$, we have chosen simple structural metrics that work for both functions and methods (\ie LOC, cyclomatic complexity, number of comments), as well as warnings from popular static analysis tools for Python and Java. It is possible that other tools might produce results different from what we have observed.
\rev{Regarding RQ$_2$, our metric selection covers the complexity and size dimensions identified by Nunez-Varela~\etal~\cite{nunez2017source} but does not exhaustively address all quality dimensions (\eg coupling, cohesion, test coverage). It is possible that complementary metrics or alternative static analysis tools might reveal additional quality differences not captured by our current analysis.}
\rev{Additionally, our study uses \gpt rather than the full GPT-4o model. While \gpt represents a significant advancement over GPT-3.5 and has demonstrated competitive performance in code generation tasks, it may not fully reflect the capabilities of the larger GPT-4o variant. The choice was driven by practical cost considerations given the scale of our experimental design (11,500 API calls per model). Future work could replicate the study with the full GPT-4o model to assess whether the observed patterns hold with the higher-capacity variant.}
\rev{Furthermore, our evaluation uses only the method signature and docstring as input, without incorporating repository-level context such as cross-file dependencies or surrounding code. While this ensures a controlled comparison across prompt languages, it may not fully reflect real-world development scenarios where LLMs have access to broader project context. Future work could investigate whether providing repository-level context influences the observed language-dependent patterns.} \revminor{Building on this point, a specific fairness consideration arises because repository-level artifacts in the studied benchmarks are predominantly written in English. Our findings, therefore, characterize prompt-language effects in isolation from any English-language signal that repository context would carry, and they do not indicate whether the fairness gaps observed for non-English prompts would persist, shrink, or widen once such context is factored in.}

Threats to \textbf{\emph{internal validity}} concern factors internal to our study that could affect our results. One such factor is the choice of the prompts. We have chosen guidelines from previous work about code generation \cite{shinn2023advances}. Concerning the translations, it is possible that the way native speakers produced the translation might have influenced the results. However, all followed guidelines to avoid enriching the prompt beyond a literal translation, for which they were driven by an LLM-generated one.
\rev{While the two-stage process (GPT-4 draft followed by expert manual curation) provides an inherent form of cross-validation, having multiple independent translators per language would further strengthen translation reliability. We acknowledge this as a limitation of the current study and encourage future work to incorporate independent multi-translator vetting for each target language.}
Another factor is the LLM  non-determinism, which we mitigated by (i) setting the temperature to zero and the seed to a constant value and (ii) repeating the queries ten times. 
For RQ$_3$, we could not observe differences in the source code identifiers because they were biased by the signature in the prompt required to execute test cases. As shown in \secref{sub:rq3}, we mitigated this threat by performing an experiment on entirely translated (in Italian) prompts, limited only to RQ$_3$, because such code could not be integrated and tested with the rest of its context. The performed study 
 showed how a prompt in Italian led to identifiers being mostly translated. Note that, as pointed out in \secref{sec:rq3method}, the' nationalization of identifiers is less likely to be performed for languages that do not use Latin characters. Clearly, it is not feasible to perform the whole study (including RQ$_1$ and RQ$_2$ on fully-translated queries, because, as explained in \secref{sec:rq3method}, such code could not be integrated with its context (\eg the class/file to which it belongs), nor with its test cases.
\rev{We acknowledge that the controlled experiment on translated signatures was conducted only on Python. Given the notably different inconsistency patterns observed between Python and Java in Table~10, extending this ablation to Java would provide additional insights into how the programming language interacts with prompt language in determining lexicon consistency. We identify this as a direction for future work.}
\noindent Finally, we cannot fully prove the cause-and-effect relationship between prompts' language and code quality, as this would require a deep knowledge of the models' architectures and, above all, training data, which are not disclosed.

Threats to \textbf{\emph{conclusion validity}} concern the relationship between experimentation and outcome. %Wherever possible, and in particular for RQ$_1$ and RQ$_2$, 
Where appropriate, we used suitable statistical procedures, $p$-value adjustments, and effect size measures to support our claims. 
%However, as remarked before, given the goals of the study and the artifacts being considered, the effect size magnitude is not particularly important to consider, except for some metrics.

Threats to \textbf{\emph{external validity}} concern the generalizability of our findings.
Such a threat, first of all, concerns the choice of the dataset (CoderEval), which, however, is a state-of-the-art benchmark for code generation tasks. \rev{To mitigate this threat, we replicate our findings on a second benchmark, ClassEval (class-level, Python only), which demonstrates that the core findings hold across different task granularities. While the addition of ClassEval strengthens generalizability, further replication on additional benchmarks -- particularly for Java -- remains desirable. This is particularly true when the model is provided with a task whose complexity extends the reach of single code units to multiple files.} Secondly, it may affect the choice of LLMs. We used state-of-the-art LLMs accessible through APIs. \rev{Our selection criterion was API accessibility, which ensures consistency in the experimental setup, reproducibility of the generation process, and uniform control over parameters such as temperature and seed. We note that DeepSeek, while accessed via API in our study, also has open-source versions available. Future work could consider locally deployed open-source models to assess whether hosting configuration influences the observed patterns.}   We considered three of the most widely used languages worldwide (plus Italian), yet the study could be extended to other widely adopted languages, such as French or Arabic. \rev{Our language selection was driven by linguistic diversity across writing systems (logographic, Devanagari, Latin-based) and global prominence, rather than by training data representation. While all four selected languages can be considered high-resource, extending the study to low-resource languages would provide additional insights into whether LLM code generation performance is strictly tied to the volume of language-specific training data. However, our reliance on manually curated translations validated by native-speaker domain experts imposed practical constraints on the number of languages we could include. We identify the inclusion of low-resource languages as an important direction for future work.} Last but not least, given the need for running tests, using widely adopted static analysis tools, and the space available, we limited the choice of programming languages to Python and Java. Also in this case, especially given the observed differences, further studies on other languages are desirable.

% !TEX root = main.tex
\section{Conclusion and Future Work} 
\label{sec:conclusion}

In this paper, we studied the influence of language used to prompt LLM when generating source code elements on code quality, measured in terms of correctness (test success/fail), metrics, and static analysis tool warnings, as well as the language used to write comments and string literals. The study is based on 230 Python tasks and 230 Java tasks from the CoderEval~\cite{hao2024codereval} benchmark \rev{and 100 Python classes from the ClassEval~\cite{du2023classeval} benchmark}, which have been translated into Chinese, Hindi, Spanish, and Italian. We experimented with three state-of-the-art LLMs, \gpt, \ds, and \cl.

%The results of the study indicate that (i) the source code correctness does not necessarily worsen when using languages other than English, (ii) by looking at metrics and static analysis warnings, the code quality does not deteriorate, yet the generated code may look different, and (iii)  LLMs tend to generate source code comments and string literals inconsistently.
The results of the study indicate that (i) the source code correctness does not necessarily worsen when using languages other than English, (ii) \revminor{across the size, complexity, and comment metrics and the static-analysis warnings from \flake, \pylint, \pmd, and \sonar examined in this study---and, for Java, on the subset of outputs that successfully compile---the code quality does not deteriorate}, yet the generated code may look different, and (iii) LLMs tend to generate source code comments and string literals inconsistently.
\begin{comment}

\begin{compactenum}
\item The source code correctness, measured by executing the benchmark's test cases, is not necessarily worse when using languages other than English, and this is especially true for Python and Chinese.
\item The source code quality indicators do not tell that code produced with prompts in languages is worse than when using English, although they suggest that it is different, \eg longer, shorter, with more or fewer decision points, and with different comment lengths.
\item LLMs tend to generate source code comments and string literals inconsistently.
\end{compactenum}
\end{comment}

Work-in-progress aims to extend the study through several dimensions. These include enlarging the set of considered languages (as well as programming languages) and performing further, deeper analyses on the quality differences for the generated code, for instance, concerning readability and identifier quality, among other quality attributes. Also, we could use explainability techniques to investigate the extent to which input in given languages may lead to different code generation. \rev{In particular, extending the study to low-resource languages (\eg Swahili, Bengali, or Yoruba) would enable a more rigorous assessment of whether LLM code generation capabilities generalize across diverse linguistic contexts or remain tied to the volume of language-specific training data.} \revminor{A further promising direction is the design of retrieval-augmented pipelines that translate repository-level context (\eg surrounding code and existing comments) into the prompt language, allowing us to disentangle prompt-language effects from the influence of English-only repository artifacts and to assess whether such configurations further improve fairness for non-English prompts.}

% !TEX root = main.tex
\section{Data Availability}
\label{sec:da}
Datasets, scripts, and raw results used for conducting the study are available in our replication package \cite{replication}.

\bibliographystyle{ACM-Reference-Format}
\bibliography{main}
\end{document}